\documentclass[useAMS,usenatbib]{mnras}
\usepackage{graphicx}
\usepackage{amsmath, amssymb}
\usepackage[T1]{fontenc}
\usepackage[usenames]{color}
\usepackage[dvipsnames]{xcolor}
\usepackage{bm}
\usepackage[capitalize]{cleveref}
\usepackage{physics}
\usepackage{acro}
\usepackage{caption}
\usepackage{subcaption}
\usepackage{stfloats}
\usepackage{xspace}
\usepackage{hyperref}
\usepackage{booktabs}
\usepackage{tabularx}
\usepackage{multicol}
\usepackage{ulem}
\usepackage{url}
\usepackage{orcidlink}
\usepackage{siunitx}

\newcommand{\zobs}{z_{\rm obs}}
\newcommand{\zpred}{z_{\rm pred}}
\newcommand{\ztrue}{z_{\rm true}}
\newcommand{\zCMB}{z_{\rm obs}}
\newcommand{\zcosmo}{z_{\rm cosmo}}
\newcommand{\zpec}{z_{\rm pec}}

\newcommand{\Vpec}{V_{\rm pec}}

\newcommand{\Mpch}{\ensuremath{h^{-1}\,\mathrm{Mpc}}}
\newcommand{\Msunh}{\ensuremath{h^{-1}\,\mathrm{M}_\odot}}
\newcommand{\kmsec}{\ensuremath{\mathrm{km}\,\mathrm{s}^{-1}}}
\newcommand{\kmsecMpc}{\ensuremath{\;\mathrm{km}\,\mathrm{s}^{-1}\,\mathrm{Mpc}^{-1}}}

\newcommand{\mpred}{m_{\rm pred}}
\newcommand{\mobs}{m_{\rm obs}}
\newcommand{\etaobs}{\eta_{\rm obs}}
\newcommand{\etatrue}{\eta_{\rm true}}

\newcommand{\xonetrue}{x_{1,\mathrm{true}}}
\newcommand{\xoneobs}{x_{1,\mathrm{obs}}}
\newcommand{\cobs}{c_{\mathrm{obs}}}
\newcommand{\ctrue}{c_{\mathrm{true}}}

\newcommand{\sint}{\sigma_{\rm int}}

\newcommand{\sigmaTFR}{\sigma_{\rm TFR}}
\newcommand{\aTFR}{a_{\rm TFR}}
\newcommand{\bTFR}{b_{\rm TFR}}
\newcommand{\cTFR}{c_{\rm TFR}}

\newcommand{\MSN}{M_{\rm SN}}

\newcommand{\Om}{\Omega_{\rm m}}

\newcommand{\SFI}{SFI\texttt{++}\xspace}
\newcommand{\TWOMPP}{2M\texttt{++}\xspace }

\defcitealias{Carrick_2015}{C15}
\defcitealias{Lilow2024}{L24}
\defcitealias{Sorce_2018}{S18}
\defcitealias{Courtois2023}{C23}

\newcommand{\CBa}{\textcolor{blue}{\texttt{CSiBORG1}}\xspace}
\newcommand{\CBb}{\textcolor{blue}{\texttt{CSiBORG2}}\xspace}

\DeclareAcronym{CMB}{short = CMB, long  = cosmic microwave background}
\DeclareAcronym{LOS}{short = LOS, long  = line-of-sight}
\DeclareAcronym{IC}{short = IC, long  = initial condition}
\DeclareAcronym{BORG}{short = \texttt{BORG}, long  = \textit{Bayesian Origin Reconstruction from Galaxies}}
\DeclareAcronym{CSIBORG}{short = \texttt{CSiBORG}, long  = \textit{Constrained Simulations in} \texttt{BORG}}
\DeclareAcronym{BIC}{short = BIC, long  = Bayesian information criterion}
\DeclareAcronym{TFR}{short = TFR, long  = Tully--Fisher relation}
\DeclareAcronym{LCDM}{short = $\Lambda$CDM, long  = $\Lambda$-cold dark matter}
\DeclareAcronym{CF4}{short = CF4, long  = CosmicFlows-4}
\DeclareAcronym{CF2}{short = CF2, long  = CosmicFlows-2}
\DeclareAcronym{HMC}{short = HMC, long  = Hamiltonian Monte Carlo}
\DeclareAcronym{MCMC}{short = MCMC, long  = Markov chain Monte Carlo}
\DeclareAcronym{LG}{short= LG, long = Local Group}
\DeclareAcronym{SPH}{short = SPH, long = smoothed-particle hydrodynamics}
\DeclareAcronym{SDSS}{short = SDSS, long = Sloan Digital Sky Survey}
\DeclareAcronym{WISE}{short = WISE, long = Wide-field Infrared Survey Explorer}
\DeclareAcronym{SN}{short = SN, long  = supernova, short-plural = e, long-plural  = e}

\title[The Velocity Field Olympics]{The Velocity Field Olympics: Assessing velocity field reconstructions with direct distance tracers}

\author[R. Stiskalek et al.]{Richard Stiskalek,$^{1,2}$\thanks{\href{mailto:richard.stiskalek@physics.ox.ac.uk}{richard.stiskalek@physics.ox.ac.uk}}\orcidlink{0000-0002-0986-314X}
Harry Desmond,$^{3}$\orcidlink{0000-0003-0685-9791}
Julien Devriendt,$^{1}$\orcidlink{0000-0002-8140-0422}
Adrianne Slyz,$^{1}$\orcidlink{0000-0002-9613-9044}\newauthor
Guilhem Lavaux,$^{4}$\orcidlink{0000-0003-0143-8891}
Michael J. Hudson,$^{5,6,7}$\orcidlink{0000-0002-1437-3786}
Deaglan J. Bartlett,$^{4}$\orcidlink{0000-0001-9426-7723}\newauthor
H\'el\`ene M. Courtois$^{8}$\orcidlink{0000-0003-0509-1776}
\\
$^{1}$Astrophysics, University of Oxford, Denys Wilkinson Building, Keble Road, Oxford, OX1 3RH, UK\\
$^{2}$Center for Computational Astrophysics, Flatiron Institute, 162 5th Ave, New York, NY 10010, USA\\
$^{3}$Institute of Cosmology \& Gravitation, University of Portsmouth, Dennis Sciama Building, Portsmouth, PO1 3FX, UK\\
$^{4}$CNRS \& Sorbonne Universit\'e, Institut d'Astrophysique de Paris (IAP), UMR 7095, 98 bis bd Arago, F-75014 Paris, France\\
$^{5}$Waterloo Centre for Astrophysics, University of Waterloo, Waterloo, ON N2L 3G1, Canada\\
$^{6}$Department of Physics and Astronomy, University of Waterloo, Waterloo, ON N2L 3G1, Canada\\
$^{7}$Perimeter Institute for Theoretical Physics, Waterloo, ON N2L 2Y5, Canada\\
$^{8}$Universit\'e Claude Bernard Lyon 1, IUF, IP2I Lyon, 69622 Villeurbanne, France\\
}

\date{Accepted XXX. Received YYY; in original form ZZZ}
\pubyear{2025}

\begin{document}\label{firstpage}
\pagerange{\pageref{firstpage}--\pageref{lastpage}}
\maketitle

\begin{abstract}
The peculiar velocity field of the local Universe provides direct insights into its matter distribution and the underlying theory of gravity, and is essential in cosmological analyses for modelling deviations from the Hubble flow. Numerous methods have been developed to reconstruct the density and velocity fields at $z \lesssim 0.05$, typically constrained by redshift-space galaxy positions or by direct distance tracers such as the Tully--Fisher relation, the fundamental plane, or Type Ia supernovae. We introduce a validation framework to evaluate the accuracy of these reconstructions against catalogues of direct distance tracers. Our framework assesses the goodness-of-fit of each reconstruction using Bayesian evidence, residual redshift discrepancies, velocity scaling, and the need for external bulk flows. Applying this framework to a suite of reconstructions---including those derived from the \textit{Bayesian Origin Reconstruction from Galaxies} (\texttt{BORG}) algorithm and from linear theory---we find that the non-linear \texttt{BORG} reconstruction consistently outperforms others. We highlight the utility of such a comparative approach for supernova or gravitational wave cosmological studies, where selecting an optimal peculiar velocity model is essential. Additionally, we present calibrated bulk flow curves predicted by the reconstructions and perform a density--velocity cross-correlation using a linear theory reconstruction to constrain the growth factor, yielding $S_8 = 0.793 \pm 0.035$. The result is in good agreement with both weak lensing and \textit{Planck}, but is in strong disagreement with some peculiar velocity studies.
\end{abstract}

\begin{keywords}
large-scale structure of the universe -- galaxies: distances and redshifts -- cosmology: distance scale
\end{keywords}


\section{Introduction}\label{sec:intro}

Spectroscopic analysis of the sky, particularly of galaxies, provides fundamental insights into the nature of our Universe. The spectroscopic data is primarily encoded in redshift measurements, which combine two key components: the uniform cosmic expansion and the peculiar motions of galaxies relative to this expansion. These motions can be characterized by the first-order moment of the phase-space distribution, known as the peculiar velocity field.

The peculiar velocity field, representing the time-integrated acceleration of galaxies since their formation, serves as a powerful probe of gravitational dynamics. It plays a crucial role in validating cosmological models by enabling direct comparisons between observed galactic motions and theoretical predictions based on the dark matter distribution~\citep{Yahil_1991,PikeHudson:2005,Davis_2011,Carrick_2015,AdamsBlake:2017,Dupuy:2019}. Furthermore, this field provides valuable insights into structure formation processes and helps constrain fundamental cosmological parameters~\citep{Nusser_2017,Huterer_2017,Boruah_2019,AdamsBlake:2020,Said_2020}.

The importance of peculiar velocity fields extends to major contemporary astronomical initiatives. Programmes such as SH0ES~\citep{Riess:2011,Riess:2022} and LIGO--Virgo--KAGRA~\citep{LVK,Abbott:2021} incorporate these measurements to account for peculiar velocities in their determinations of the Hubble constant. The peculiar velocities for the former were derived by~\cite{Peterson_2022,Carr_2022}, primarily based on the reconstruction by~\citealt{Carrick_2015}, whereas the latter relies on compilations of peculiar velocities, such as GLADE\texttt{+}~\citep{Dalya:2022}, which is based on~\citealt{Jasche_2019}. This application is particularly relevant in the context of current cosmological tensions, as peculiar velocity measurements can shed light on the $S_8$ tension through their relationship with the growth rate parameter $f \sigma_8 \sim S_8$~\citep{KIDS1000,DES_Y3,Amon:2022,Madhavacheril:2024}.

Various methodologies have been developed to reconstruct the velocity field, utilizing either redshift-space galaxy positions or direct peculiar velocity measurements. However, these reconstructions must carefully account for selection effects to avoid systematic biases. These methodological challenges, particularly in the context of comparing density and peculiar velocity fields of nearby galaxies, have been extensively analysed in the literature~\citep{StraussWillick_1995}.

Several previous works have developed increasingly complex frameworks to compare velocity models to distance data fairly \citep{Boruah_2019, Said_2020}. In this work, we introduce a self-consistent framework to assess the accuracy of current reconstruction models available in the scientific literature. We account for the homogeneous and inhomogeneous Malmquist biases \citep{StraussWillick_1995}, distance tracer selection effects, and joint inference of standardization parameters. We apply this framework to a variety of existing models from the literature and data pairs. This includes the \ac{CSIBORG} models based on the \acl{BORG} algorithm (\ac{BORG}, e.g.~\citealt{Jasche_2019}), along with other models from the literature \citep{Carrick_2015,Sorce_2018,Sorce_2020,Lilow2024}. We aim to determine which models and methods are most consistent with distance data and to understand the systematic biases in each model. This should serve as a guideline for future studies that aim to use peculiar velocity data to constrain cosmological parameters and help them choose the best model for their analysis.

The article is structured as follows. In~\cref{sec:data}, we describe the peculiar velocity field models derived from reconstructions and distance data used in this work. In~\cref{sec:method}, we introduce our statistical and validation framework. In~\cref{sec:results}, we present the results of our analysis, which include the comparison of the evidence for the different velocity models and consistency parameters with cosmology. In~\cref{sec:discussion}, we discuss the implications of our findings. Finally, in~\cref{sec:conclusion}, we summarise our conclusions. We note that all logarithms in this work are base-10, unless explicitly stated. We use the notation $\mathcal{N}(x; \mu, \sigma)$ to denote the normal distribution with mean $\mu$ and standard deviation $\sigma$ evaluated at $x$.


\section{Data}\label{sec:data}

In this section, we discuss first the different kinds of velocity model reconstructions we use in this work (\cref{sec:reconstructions}), and then the distance data we use to validate these models (\cref{sec:peculiar_velocity_samples}).

\subsection{Local Universe reconstructions}\label{sec:reconstructions}

We use several reconstructions of the local Universe, constrained by either redshift-space galaxy positions or peculiar velocities. The former approach typically employs a counts-per-voxel likelihood and requires a galaxy bias to predict the number of galaxies in a cell based on the cell's matter density, after which it is limited mainly by shot noise. The latter relies on a (Gaussian) likelihood defined in the projected peculiar velocity, but is limited by the large noise in peculiar velocity measurements. We summarise these reconstructions in~\cref{tab:reconstructions}, and describe them one-by-one here.

\subsubsection{Carrick et al.}

In~\citet[hereafter~\citetalias{Carrick_2015}]{Carrick_2015}, the luminosity-weighted density field is derived from the redshift-space positions of galaxies in the \TWOMPP\ catalogue, using the iterative scheme of~\cite{Yahil_1991}. \TWOMPP~is a whole-sky redshift compilation of $69,160$ galaxies~\citep{Lavaux_2011}, derived from photometry from the Two-Micron-All-Sky Extended Source Catalog~\citep{Skrutskie_2006} and redshifts from the 2MASS Redshift Survey~\citep[2MRS,][]{Huchra_2012}, the 6dF Galaxy Redshift Survey~\citep{Jones_2009}, and the \ac{SDSS} Data Release 7~\citep{Abazajian_2009}. The \TWOMPP\ $K$-band apparent magnitudes are corrected for Galactic extinction, $k$-corrections, evolution, and surface brightness dimming. The catalogue is apparent magnitude limited to $K < 11.5$ in the regions of sky covered by 2MRS and to $K < 12.5$ within the areas covered by 6dF and \ac{SDSS}.

\citetalias{Carrick_2015} first fits a luminosity function to the \TWOMPP\ catalogue to assign luminosity weights to each galaxy. Then, the luminosity-dependent bias of~\cite{westover} is applied to normalise the density contrast across all radii. Galaxy positions are moved from the redshift-space to real-space using an iterative approach. Within the inner region of approximately $125~\Mpch$ the Zone of Avoidance is masked in the \TWOMPP\ data and filled by ``cloning'' galaxies above and below the Galactic plane to reconstruct the density field in this region. The resulting velocity field is calculated from the density field assuming linear perturbation theory. The velocity field is independent of the dimensionless growth factor, and thus must be scaled to match the peculiar velocities (with a factor $\beta^\star$ in the notation of~\citetalias{Carrick_2015}). This single field is generated on a $256^3$ grid with a box size of $400~\Mpch$, a voxel size of $1.6~\Mpch$, and assuming $\Om = 0.3$.

The parameter $\beta^\star$ is defined by
\begin{equation}\label{eq:beta_star}
    \beta^\star \equiv \frac{f\sigma_{8,\mathrm{NL}}}{\sigma_8^b}
\end{equation}
where $f$ is the dimensionless growth rate defined as $f \equiv \dd \ln D / \dd \ln a$, with $D$ representing the growth function of linear perturbations and $a$ being the scale factor. In \ac{LCDM}, $f \approx \Om^{0.55}$~\citep{Bouchet_1995,Wang_1998}, though in modified gravity theories this index can differ from $0.55$ (e.g.~\citealt{Dvali_2000, Linder_2007}). The term $\sigma_8^b$ refers to the typical fluctuation scale in the galaxy over-density field at a radius of $8~\Mpch$ (the `b' denotes that it is measured from a biased tracer, galaxies). $\sigma_{8,\mathrm{NL}}$ is the typical fluctuation scale of the non-linear matter field at a radius of $8~\Mpch$. The value of $\sigma_8^b$ in the \TWOMPP\ data has been measured by~\cite{westover} to $0.98 \pm 0.07$ by fitting projected correlation functions to 2MRS galaxies~\citep{Huchra_2012}, whereas~\citetalias{Carrick_2015} measured $\sigma_8^b = 0.99 \pm 0.04$ by using the maximum-likelihood scheme of counts in cells within radial shells~\citep{Efstathiou_1990}. Thus,~\citetalias{Carrick_2015} together with peculiar velocity samples can be used to constrain the growth of structure or the $S_8$ parameter (see e.g.~\citealt{Boruah_2019, Said_2020}). The value of $\beta^\star$ in case of~\citetalias{Carrick_2015} calibrated against LOSS, Foundation, 2MTF, and~\SFI~has been previously studied extensively in~\cite{Boruah_2019}.

\subsubsection{Lilow et al.}

In~\citet[hereafter~\citetalias{Lilow2024}]{Lilow2024}, the density and peculiar velocity fields are reconstructed using the 2MRS data, but with a machine learning approach, rather than linear theory. \citetalias{Lilow2024} employ an autoencoder with a U-Net architecture, previously tested in their earlier work~\citep{VeenaGaneshaiah_2023}. The network was trained separately for density and velocity on \texttt{Quijote}-based $N$-body simulation mocks~\citep{Quijote} including redshift-space distortions, galaxy bias, and the 2MRS selection function, with the velocity field constrained to be irrotational. Comparison to Wiener filter reconstructions on mock data showed substantially lower reconstruction errors and an accurate recovery of the non-linear velocity–density relation. Notably, the network captures information on large-scale bulk flows through redshift-space distortions, and when applied to 2MRS it recovers known clusters and a Local Group velocity closely aligned with the observed CMB dipole. The trained model is subsequently applied to the 2MRS data. \citetalias{Lilow2024} assume cosmological parameters from the fiducial \texttt{Quijote} simulations: $\Omega_m = 0.3175,~\Omega_{\rm b} = 0.049,~h=0.6711,~n_s = 0.9624,~\sigma_8 = 0.834$. The single field is presented on a $128^3$ grid with a box size of $400~\Mpch$ and a voxel size of $3.1~\Mpch$.

\subsubsection{\texttt{CSiBORG}}\label{sec:csiborg}

The \ac{CSIBORG} suite is a set of cosmological $N$-body simulations that are constrained to match the 3D mass distribution of the local Universe. This is achieved by imposing~\acp{IC} on the density field derived from the \ac{BORG} algorithm applied to~\TWOMPP. \ac{BORG} produces a posterior of voxel-by-voxel densities at $z=1000$ by applying a gravity forward model, redshift-space distortions, observational selection effects and a biasing prescription to the~\acp{IC} before comparing to the galaxy number density field of~\TWOMPP with a Poisson likelihood~\citep{Jasche_2013,Lavaux_2016,Jasche_2019,Lavaux_2019,Porqueres_2019}. Each \ac{CSIBORG} box resimulates a single \ac{BORG} posterior sample, and thus the variation across the full set permits quantification of the uncertainty in the dark matter field due to incomplete knowledge of the galaxy field and its relationship to the underlying density.

Here we employ two versions of the \ac{CSIBORG} simulations, based on different versions of the \ac{BORG} \TWOMPP\ \acp{IC}. The first suite, \CBa, uses the \ac{BORG} \TWOMPP\ \acp{IC} from~\cite{Jasche_2019}, while the second suite, \CBb, is based on the updated \ac{BORG} \TWOMPP\ \acp{IC} from~\cite{Stopyra_2023}. Both versions of the \acp{IC} are constrained within the \TWOMPP\ region, covering a spherical volume with a radius of approximately $155~\Mpch$ centred on the Milky Way ($z \lesssim 0.06$), embedded in a box of $1000~\mathrm{Mpc}$. The regions outside the \TWOMPP\ volume are largely unconstrained due to catalogue selection
providing the expected long-range forces in a \ac{LCDM} framework. The \ac{BORG} algorithm models the density field on a $256^3$ grid, yielding a spatial resolution of $2.65~\Mpch$.

The \CBa\ suite, first introduced in~\cite{Bartlett_2021}, consists of $101$ dark matter-only $N$-body simulations in a $677.7~\Mpch$ box centred on the Milky Way. The constrained \acp{IC} from \ac{BORG} are linearly evolved to $z = 69$ and augmented with white noise within the constrained \TWOMPP\ region on a $2,048^3$ grid, achieving a spatial resolution of $0.33~\Mpch$ and a dark matter particle mass of $3 \times 10^9\Msunh$. A buffer region of around $10~\Mpch$ surrounds the high-resolution region, ensuring a smooth transition to the unconstrained, low-resolution outer region. For this work, however, we use only $20$ of these resimulations to reduce the computational cost, though we verify that our results are not significantly affected by using all $101$ resimulations. The initial conditions are evolved to $z = 0$ using the adaptive mesh refinement code \texttt{RAMSES}~\citep{Teyssier2002_RAMSES}. The \CBa\ suite adopts cosmological parameters from~\cite{Planck2014}, with $H_0$ set according to the 5-year WMAP results combined with Type Ia \ac{SN} and baryon acoustic oscillation data~\citep{Hinshaw2009}: $\Omega_m = 0.307,~\sigma_8=0.8288,~H_0 = 70.5~\kmsecMpc,~n_s = 0.9611,~\Omega_{\rm b} = 0.04825$.

The \CBb\ suite follows a similar setup but uses newer \ac{BORG} \acp{IC}~\citep{Stopyra_2023}, which use a more accurate gravity model: a $20$-step \texttt{COLA} integrator~\citep{Tassev_2013_COLA} instead of the $10$-step particle mesh solver of \CBa. This was shown to improve the agreement of cluster masses with observational data. The cosmological parameters for \CBb\ are based on~\citep{Planck_2020}, with lensing and baryon acoustic oscillations: $\Omega_m = 0.3111,~\sigma_8 = 0.8102,~H_0 = 67.66~\kmsecMpc,~n_s = 0.9665,~\Omega_{\rm b} = 0.049$. We resimulate $20$ samples of the \ac{BORG} posterior ourselves using \texttt{Gadget4}~\citep{Gadget4}. We note that~\cite{Stopyra_2023} already resimulated $20$ samples of the \ac{BORG} posterior, but without the zoom-in technique: the~\ac{CSIBORG} setup allows us to reach $\sim$200 times better mass resolution in the constrained region. In both \ac{CSIBORG} suites, we construct the $z = 0$ density and velocity fields from the particle snapshots using the \ac{SPH} technique~\citep{Monaghan_1992,Colombi_2007} and set the minimum number of points to smooth over to $32$ (following Section IV.B.1 of~\citealt{Bartlett_2022}). We render the final fields in voxels of $0.7~\Mpch$.

Unlike~\citetalias{Carrick_2015}, which is derived independently of the growth factor and consequently of $\sigma_8$, the derivations of the \ac{BORG}-based reconstructions (and all others presented in this work) assume fixed values for $\Om$ and $\sigma_8$. They naturally account for the $\beta^\star$ factor of~\cref{eq:beta_star}, which is effectively pinned to 1. However, to search for systematics in the reconstructions leading to a mismatch with the peculiar velocity data, we allow the velocity fields other than~\citetalias{Carrick_2015} to be scaled by a free parameter $\beta$, with the fiducial value of 1, similarly to the $A_L$ parameter in \ac{CMB} lensing~\citep{Calabrese_2008}. We retain the notation of $\beta^\star$ for the scaling factor of~\citetalias{Carrick_2015}, who measured $\beta^\star = 0.431 \pm 0.021$.

\subsubsection{Sorce}

The reconstructions presented in~\citet[hereafter~\citetalias{Sorce_2018}]{Sorce_2018,Sorce_2020} employ the technique of constrained realisations. Unlike redshift space density–constrained methods, these simulations use only radial peculiar velocities---drawn from the grouped \ac{CF2} catalogue~\citep{CosmicFlows2} of $\sim$\num{5000} objects---as constraints. Grouping suppresses virial motions, while a bias–minimisation scheme iteratively restores Gaussianity to the velocity distribution, reducing spurious infall. The cosmic displacement field is reconstructed with a Wiener filter and constraints are relocated to progenitor positions using the reverse Zel'dovich approximation, with noisy radial velocities replaced by their Wiener–filtered 3D estimates. Constrained realisations then combine these fields with Gaussian random modes to statistically restore missing large- and small-scale power.

\citetalias{Sorce_2018} thus produces \acp{IC} constrained solely by peculiar velocities, which are evolved to $z=0$ with \texttt{Gadget2}~\citep{Gadget2} in a $500~\Mpch$ box ($512^3$ particles with a particle mass of $8\times10^{10}\Msunh$). Subsequent refinements introduced improved modelling of peculiar–velocity uncertainties, enabling stable simulacra of local clusters (Virgo, Centaurus, Hydra, Coma).~\cite{Sorce_2020} incorporated the ``paired fixed field'' technique~\citep{Pontzen_2016}, demonstrating that about 80 per cent of the large–scale power is already constrained by local data and provided a method to estimate uncertainties without hundreds of realisations. We use a single such simulation, assuming Planck cosmology~\citep{Planck2014}, and construct \ac{SPH} density and velocity fields on a $0.7~\Mpch$ grid.

\subsubsection{Courtois et al.}

In~\citet[hereafter~\citetalias{Courtois2023}]{Courtois2023}, the density and velocity fields of the local Universe below $\zCMB = 0.08$ are reconstructed using the \acl{CF4} catalogue (\ac{CF4};~\citealt{Tully_2023}), which includes distances and redshifts for $38,000$ groups. Similarly to~\citetalias{Sorce_2018}, the reconstruction is constrained by peculiar velocities and employs a forward-modelling procedure with a \ac{HMC} algorithm~\citep{Graziani_2019}.

Given the reported \ac{CF4} redshifts and distances, they sample the ``true'' distance to each source, the density field along with cosmological and nuisance terms such as the growth rate or the non-linear velocity dispersion $\sigma_{\rm NL}$, while also marginalising over a zero-point parameter $h_{\rm eff}$ to absorb calibration differences between distance indicators. Under the assumption of linear theory, the velocity field is derived from the density field and compared with the observed peculiar velocities. The reconstruction is performed on grids of resolution from $64^3$ to $256^3$. We use the grouped \ac{CF4} version and adopt $100$ posterior samples at $256^3$ resolution. From these reconstructions, they obtain a growth rate $f\sigma_8 = 0.36\pm0.05$ (grouped \ac{CF4}). The implied bulk flow is $230 \pm 136~\kmsec$ at $300~\Mpch$, and distances are compatible with $H_0 \simeq 74.5\pm0.1~\kmsecMpc$.

\subsubsection{Visual comparison}

\begin{figure*}
    \centering
    \includegraphics[width=\textwidth]{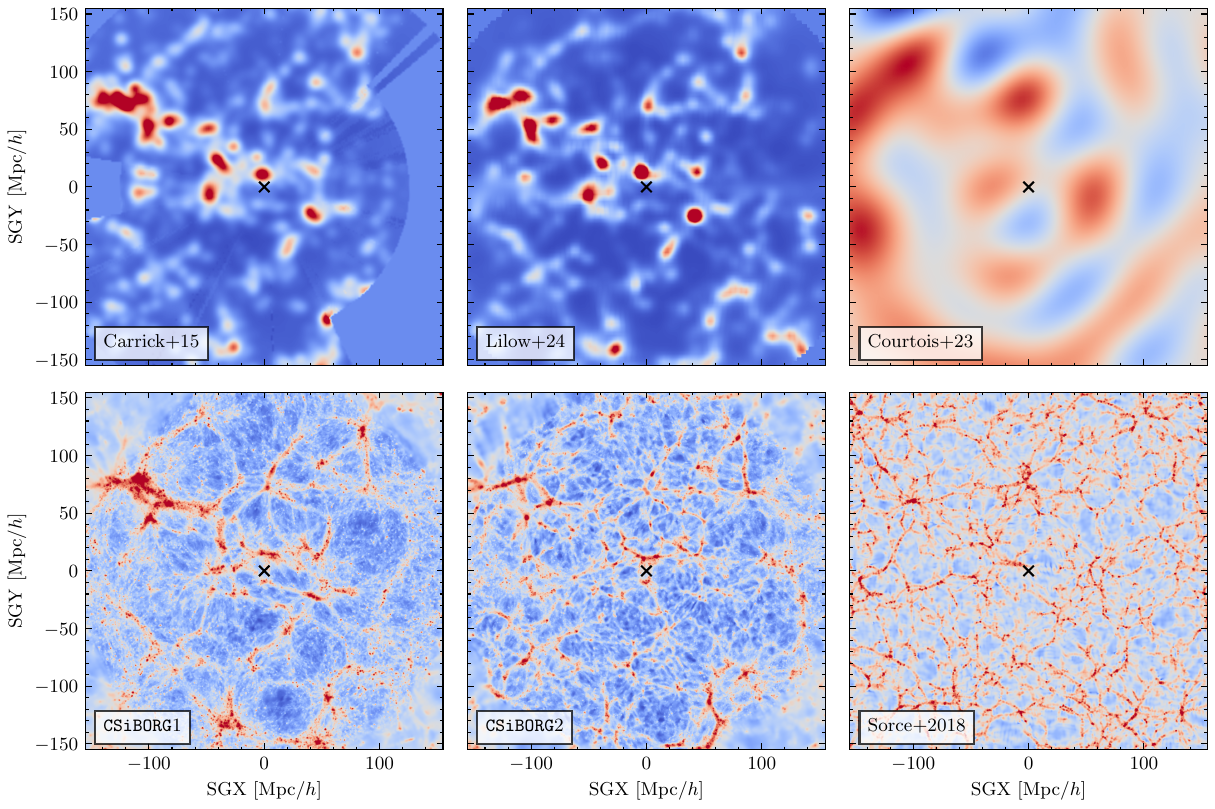}
    \caption{Slices of the density field in the $\mathrm{SGX}$-$\mathrm{SGY}$ plane from $-155~\Mpch$ to $155~\Mpch$ in supergalactic coordinates at $z = 0$ for the local Universe reconstructions used in this work (\cref{tab:reconstructions}). Redder (bluer) colours correspond to overdensities (underdensities). The density fields are presented without additional smoothing. For the fields in the bottom row we resimulated the \acp{IC} at higher resolution. The \ac{CSIBORG} suites and~\protect\cite{Courtois2023} are averaged over $20$ posterior samples. The black cross marks the origin, indicating the approximate position of the Local Group.}
    \label{fig:XY_slices}
\end{figure*}

We show slices through the density field at $z = 0$ from the six reconstructions we consider in~\cref{fig:XY_slices}.~\citetalias{Carrick_2015} and~\citetalias{Lilow2024} are qualitatively similar, but do not display significant cosmic web-like features. In contrast, the \ac{CSIBORG} fields reveal a prominent cosmic web structure due to their use of non-linear modelling. This is true even after averaging the density field over $20$ independent posterior samples, suggesting that it is not due to the unconstrained white noise of these simulations. The~\citetalias{Courtois2023} field appears more diffuse despite having a voxel resolution of $3.9~\Mpch$ (similar to~\citetalias{Carrick_2015}), which suggests that despite the voxel resolution the field varies only at much larger scales. Lastly, while the field of~\citetalias{Sorce_2018} is also constrained at the level of linear theory, it was resimulated at a much higher resolution with additional small-scale white noise, resulting in a clear cosmic web.

\begin{figure}
    \centering
    \includegraphics[width=\columnwidth]{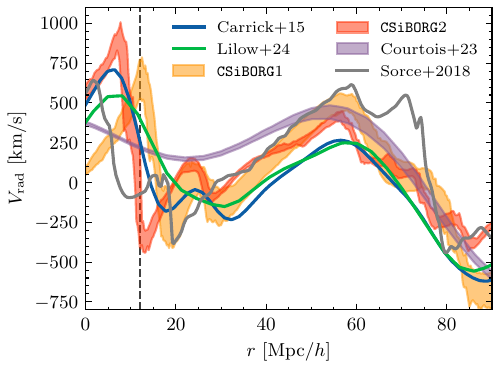}
    \caption{Examples of the reconstructed velocities along the \ac{LOS} to the Virgo cluster, whose approximate distance is indicated by the dashed vertical line. For reconstructions providing multiple samples of the velocity field, a shaded region between 16\textsuperscript{th} and 84\textsuperscript{th} percentiles is shown. For~\protect\cite{Carrick_2015} we assume the fiducial value of $\beta^\star = 0.43$.}
    \label{fig:LOS_Virgo}
\end{figure}

Unlike the \ac{CSIBORG} reconstructions, the~\citetalias{Sorce_2018} field is a single realisation and, thus, the cosmic web features are more likely due to unconstrained, small-scale modes, not matching the actual local Universe. We also show the \ac{LOS} velocity in the direction of the Virgo cluster in~\cref{fig:LOS_Virgo} as an illustration of the differences between the fields, with the bands indicating the standard deviation among the 20 realisations for the appropriate fields.

\begin{table*}
    \centering
    \begin{tabular}{llccp{6.0cm}}
        \toprule
        \textbf{Model} & \textbf{Alias} & \textbf{Constraint Data} & \textbf{Resolution} & \textbf{Description} \\
        \midrule
        \cite{Carrick_2015}   & \citetalias{Carrick_2015}   & $\mathrm{2M}\texttt{++}$ redshift-space positions   & $4~\Mpch$   & Linear inverse modelling using luminosity-weighted galaxy density field. \\
        \cite{Lilow2024}      & \citetalias{Lilow2024}      & 2MRS redshift-space positions   & $3.1~\Mpch$ & Machine learning inverse method trained on \texttt{Quijote} simulations.\\
        \cite{Jasche_2019}    & \CBa                        & $\mathrm{2M}\texttt{++}$ redshift-space positions   & $2.6~\Mpch$ & \ac{BORG} forward model; simulations from~\cite{Bartlett_2021}. \\
        \cite{Stopyra_2023}   & \CBb                        & $\mathrm{2M}\texttt{++}$ redshift-space positions   & $2.6~\Mpch$ & \ac{BORG} forward model with an improved gravity model; simulations introduced here. \\
        \cite{Sorce_2018,Sorce_2020}     & \citetalias{Sorce_2018}     & CF2 peculiar velocities                         & $1.9~\Mpch$ & Constrained realisation with Wiener filter and reverse Zel’dovich approximation; restores structures statistically from peculiar velocities. \\
        \cite{Courtois2023}   & \citetalias{Courtois2023}   & CF4 peculiar velocities                         & $3.9~\Mpch$ & \ac{HMC} sampling of the density field at $z = 0$ constrained against peculiar velocities. \\
        \bottomrule
    \end{tabular}
    \caption{Summary of local Universe reconstructions of the density and velocity fields investigated in this work.}
    \label{tab:reconstructions}
\end{table*}

\subsection{Peculiar velocity samples}\label{sec:peculiar_velocity_samples}

\begin{figure}
    \centering
    \includegraphics[width=\columnwidth]{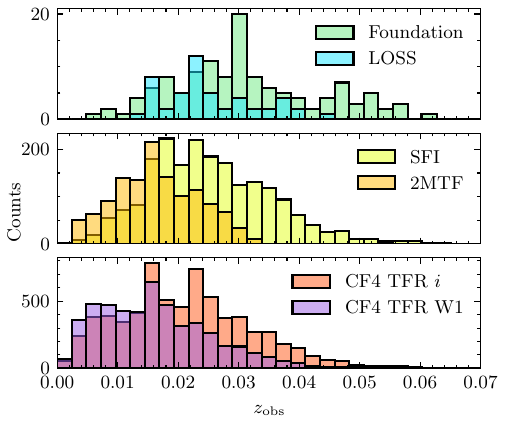}
    \caption{The distribution of observed redshifts converted to the \ac{CMB} frame ($\zCMB$) for the peculiar velocity samples used in this work. With the exception of the Foundation sample, the majority lie within $\zCMB \lesssim 0.05$, while the 2MTF sample is constrained to $\zCMB \lesssim 0.03$ due to its magnitude limit. For visual clarity, the samples are arbitrarily separated into three rows, sharing the $x$-scale but each with its own $y$-scale.}
    \label{fig:zcmb_dist}
\end{figure}

We use the \ac{TFR} and \ac{SN} samples to calibrate the flow models, which we now discuss in turn. We do not use absolute calibration of the distance tracers, so the zero-point of the distance indicators is degenerate with the Hubble constant. However, since we work in distance units of $h^{-1}$Mpc, our results are independent of $H_\mathrm{0}$ except for its influence on the power spectrum of the reconstructions. We show the samples' observed redshift distribution in the \ac{CMB} frame in~\cref{fig:zcmb_dist}. Our method for extracting luminosity distance information from the TFR and SN data is described in Sec.~\ref{sec:method}.

\subsubsection{Tully--Fisher samples}

The \ac{TFR}~\citep{Tully_1977} is an empirical relation between the width of a spectral line of spiral galaxies $W$ (typically the HI line tracing neutral hydrogen) as a measure of the galaxies' rotation velocity, and absolute magnitude $M$ as a measure of their luminosity. We write the relation as
\begin{equation}\label{eq:TFR_absmag}
    M(\eta) =
    \begin{cases}
        \aTFR + \bTFR \eta + \cTFR \eta^2 &\mathrm{if}~\eta > 0\\
        \aTFR + \bTFR \eta &\mathrm{otherwise}
    \end{cases}
\end{equation}
where $\aTFR$, $\bTFR$, and $\cTFR$ are the zero-point, slope and curvature, respectively, though the latter term is commonly neglected as the relation is fairly linear. We reparameterise the linewidth to introduce a parameter $\eta$ such that
\begin{equation}
    \eta = \log \frac{W}{\kmsec} - 2.5.
\end{equation}
Henceforth, we refer to $\eta$ as the linewidth. We model the curvature of the \ac{TFR} in high-linewidth galaxies only, which is why the curvature term applies only to galaxies with $\eta > 0$. In the subsequent analysis we jointly infer the \ac{TFR} calibration parameters along with the intrinsic scatter $\sigmaTFR$. We use three \ac{TFR} samples: 2MTF, \SFI and \ac{CF4}, which we now describe in turn.

\textbf{2MTF}: The 2MTF survey is an all-sky sample consisting of $2,062$ galaxies with \ac{TFR} data up to redshift $\zCMB \approx 0.03$, with a $K$-band apparent magnitude limit of $11.25$~\citep{Masters_2008,Hong_2019}. We utilise the 2MTF dataset compiled by~\cite{Boruah_2019}, which removes duplicates from the~\SFI~sample, uses only $K$-band magnitudes and retains only galaxies with linewidths within the range $-0.1 < \eta < 0.2$. This final sample contains $1,247$ galaxies.

\textbf{\SFI}: The~\SFI~catalogue is an all-sky sample consisting of $4,052$ galaxies and $736$ groups up to $\zCMB \lesssim 0.05$, without a strict apparent magnitude threshold~\citep{Masters_2006, Springob_2007}. Here we use the galaxy-only sample curated by~\cite{Boruah_2019} and adopt $I$-band magnitudes, though we do not apply their strict linewidth selection because we allow for \ac{TFR} curvature. The final sample contains $2,010$ objects.

\textbf{CF4}: The \ac{CF4} TFR survey consists of $9,792$ galaxies up to $\zCMB \lesssim 0.05$, without a strict apparent magnitude threshold~\citep{Kourkchi_2020B,Kourkchi_2020A}, and is a subset of the \ac{CF4} sample~\citep{Tully_2023}. We use both the \ac{SDSS} $i$-band and \ac{WISE} W1 magnitudes: while the former is limited to the \ac{SDSS} footprint, the latter covers the entire sky. We treat the two bands as separate samples, though they are not independent, as some objects have both \ac{SDSS} and \ac{WISE} photometry. We retain objects already present in either the 2MTF or~\SFI~samples, but select only galaxies with $\eta > -0.3$, galactic latitude $|b| > 7.5^\circ$ and quality flag $5$ (best). The final \ac{SDSS} $i$-band and \ac{WISE} W1 samples contain $5,027$ and $3,278$ galaxies, respectively. When analysing both samples jointly, we preferentially retain the WISE photometry for galaxies with measurements in both bands. The publicly available \ac{CF4} catalogue does not list magnitude uncertainties, thus we set them to a fiducial (conservative) value of $0.05~\mathrm{mag}$~\citep{Kourkchi_2019}.


\subsubsection{Supernova samples}

\ac{SN} samples have been widely used as distance indicators in a variety of cosmological analyses. The SALT2 model standardises the \ac{SN} light curve~\citep{SALT2}, such that the ``corrected'' apparent magnitude $m_{\rm standard}$, which relates the absolute magnitude and distance modulus in the usual way, can be written using the Tripp formula~\citep{Tripp_1998} :
\begin{equation}\label{eq:tripp}
    m_{\rm standard} = m_{\rm obs} + \mathcal{A} x_1 - \mathcal{B} c.
\end{equation}
Here, $m_{\rm obs}$ is the measured apparent magnitude and the last two terms act to standardise the \ac{SN}: $x_1$ is the stretch parameter describing the width of the light curve, and $c$ is the \ac{SN}'s colour. $m_{\rm standard}$ together with the Type Ia \ac{SN} absolute magnitude $M_{\rm SN}$ determines the distance to the \ac{SN}. We take $M_{\rm SN}$, $\mathcal{A}$ and $\mathcal{B}$ as global parameters of the fit separately for each \ac{SN} sample, which we infer jointly with the other model parameters defined below. Note that $\mathcal{A}$ and $\mathcal{B}$ are typically denoted $\alpha$ and $\beta$, but we have renamed them here to avoid confusion with variables defined later. 

We use two \ac{SN} samples in this work: the Lick Observatory Supernova Search (LOSS;~\citealt{LOSS}) and the Foundation Supernova Survey (Foundation;~\citealt{Foundation, Foundation_cosmology}). We adopt the curated samples from~\cite{Boruah_2019}, where the LOSS sample is derived by~\cite{Ganeshalingam_2013}. In this curation, \acp{SN} already present in LOSS were removed from Foundation. The final LOSS and Foundation samples contain 55 and 125 \acp{SN}, respectively. For discussion of our choice to use only these samples, and other samples that could potentially be used, see~\cref{sec:future_work}.

\begin{table*}
    \centering
    \begin{tabular}{l l r p{9cm}}
        \toprule
        \textbf{Sample} & \textbf{Type} & \textbf{Number of objects} & \textbf{Description} \\
        \midrule
        2MTF & Tully--Fisher & 1,247 & Full-sky sample curated by~\cite{Boruah_2019}, retaining only linewidth range of $-0.1 < \eta < 0.2$ and with duplicates from the~\SFI~sample removed (partially included in the \ac{CF4} sample) \\
        \SFI & Tully--Fisher & 2,010 & Full-sky sample curated by~\cite{Boruah_2019}. No strict linewidth selection applied (partially included in the \ac{CF4} sample). \\
        CF4 (\ac{SDSS} $i$-band) & Tully--Fisher & 5,027 & Part of the \ac{CF4} survey, limited to the \ac{SDSS} footprint. We retain only $\eta > -0.3$ and galactic latitude $|b| > 7.5^\circ$. \\
        CF4 (\ac{WISE} W1-band) & Tully--Fisher & 3,278 & Part of the \ac{CF4} survey with full-sky coverage (a subset of these has $i$-band photometry as well). Same selection criteria as the \ac{CF4} \ac{SDSS} $i$-band sample. \\
        Foundation & Type Ia \ac{SN} & 125 & Based on~\cite{Foundation}; curated by~\cite{Boruah_2019}. \\
        LOSS & Type Ia \ac{SN} & 55 & Based on~\cite{Ganeshalingam_2013}; curated by~\cite{Boruah_2019}. Overlap with Foundation sample removed. \\
        \bottomrule
    \end{tabular}
    \caption{Summary of the Tully--Fisher and \ac{SN} peculiar velocity samples used to test the velocity field reconstructions.}
    \label{tab:pv_samples}
\end{table*}


\section{Methodology}\label{sec:method}

In this section, we first explain the relationship between distance and redshift through peculiar velocities (\cref{sec:distance_measures}). Next, we describe the probabilistic model used to infer the calibration parameters of the distance indicators and the velocity fields (\cref{sec:probabilistic_model}). Finally, we outline the method for computing the evidence of each model (\cref{sec:evidence_calculation}).


\subsection{Distance measures}\label{sec:distance_measures}

The total redshift of a source in the \ac{CMB} frame ($\zCMB$) is
\begin{equation}\label{eq:redshift_addition}
    1 + \zCMB = \left(1 + \zcosmo\right)\left(1 + \zpec\right),
\end{equation}
where $\zcosmo$ is the redshift due to cosmic expansion and $\zpec = \Vpec /c$ is the redshift produced by the radial peculiar velocity $\Vpec$, also in the \ac{CMB} frame. The cosmological redshift is related to the source comoving distance $r$ as (e.g.,~\citealt{Hogg1999})
\begin{equation}\label{eq:redshift_to_distance}
    r(\zcosmo) = \frac{c}{H_0} \int_{0}^{\zcosmo} \frac{\dd z^\prime}{\sqrt{\Om (1 + z^\prime)^3 + 1 - \Om}},
\end{equation}
since we assume a flat \ac{LCDM} universe dominated by non-relativistic matter and dark energy only, with $\Om$ the matter density parameter. Within the single flow approximation, $\Vpec$ at the position of the source $\bm{r}$ is
\begin{equation}\label{eq:predicted_peculiar_velocity}
    \Vpec
    =
    \left(\beta \bm{v}(\bm{r}) + \bm{V}_{\rm ext}\right) \cdot \hat{\bm{r}},
\end{equation}
where $\hat{\bm{r}}$ is the unit vector along the source's \ac{LOS} and $\bm{v}(\bm{r})$ is the 3D velocity field evaluated at $\bm{r}$.

\cref{eq:predicted_peculiar_velocity} introduces two new parameters for the velocity field, $\bm{V}_{\rm ext}$ which is the ``external'' velocity of the box, and $\beta$ which is a scaling parameter. $\bm{V}_{\rm ext}$ is a constant vector accounting for the motion of the reconstruction box relative to the rest of the Universe, and is sourced by matter outside of this volume which cannot be picked up by the reconstructions themselves. $\beta$ describes two effects. First, for the linear theory inverse reconstructions of~\citetalias{Carrick_2015} where velocities are computed from the galaxy density field, then $\beta$ is expected to be $f / b$ where $f$ is the dimensionless growth factor and $b$ is the linear galaxy bias. In this case $\beta$ contains important cosmological information. For the other reconstructions the velocities are derived from the matter field rather than the galaxy field, and hence this factor is not present and $\beta$ is expected to be one. However, we keep it there as a measure of how reliably the reconstructions are matching the peculiar velocities implied by the data: if they are systematically over- or under-estimating these velocities we will find $\beta<1$ or $\beta>1$, respectively. A third parameter we introduce for each velocity field is $\sigma_v$, a Gaussian uncertainty between the field's prediction and the ``observed'' value which is assumed to be position-independent. This describes to first order the effect of small-scale motions that cannot be captured by the velocity reconstruction methods. However, we note that~\cite{Hollinger_2024} showed on~\citetalias{Carrick_2015}-like mock reconstructions that the recovered $\sigma_v$ is inflated for galaxies near the Zone of Avoidance (which is masked in our catalogues). We refer to $\{\bm{V}_{\rm ext}, \beta, \sigma_v \}$ collectively as ``calibration'' parameters of the fields.

The distance indicators typically relate an observable property to absolute magnitude $M$, which together with the apparent magnitude $m$ yields an estimate of the distance modulus $\mu$,
\begin{equation}\label{eq:distmod_definition}
    \mu = m - M.
\end{equation}
The distance modulus is a reparametrisation of the luminosity distance $d_{\rm L}$,
\begin{equation}
    \mu = 5 \log \frac{d_{\rm L}}{\mathrm{Mpc}} + 25,
\end{equation}
while $d_{\rm L}$ is related to comoving distance via
\begin{equation}
    d_{\rm L} = (1 + \zcosmo) r.
\end{equation}



\subsection{Joint flow \& distance calibration}\label{sec:probabilistic_model}

We simultaneously calibrate both the distance indicator relation and auxiliary parameters associated with the density and velocity field, which we now introduce. We either use the provided density and radial velocity fields~\citepalias[in the case of][]{Carrick_2015,Courtois2023,Lilow2024} or construct them ourselves using \ac{SPH} from particle snapshots (for \CBa, \CBb, and~\citetalias{Sorce_2018}). We evaluate these 3D fields along each tracer's \ac{LOS} up to radial distance of $200\Mpch$ with a uniform spacing of $0.5~\Mpch$, which is much higher than both the typical reconstruction resolution and distance uncertainty. We use the linear regular grid interpolator \texttt{RegularGridInterpolator} implemented in~\texttt{scipy}\footnote{\url{https://docs.scipy.org}}~\citep{scipy}.

We jointly infer the density and velocity field calibration ($\alpha_{\rm low},\,\alpha_{\rm high},\,\rho_t$ or $b_1$, and $\beta,\,\sigma_v,\,\bm{V}_{\rm ext}$), the distance indicator calibration ($\aTFR,\,\bTFR,\,\cTFR$ for the \ac{TFR} or $M_{\rm SN},\,\mathcal{A},\,\mathcal{B}$ for \acp{SN} , and $\sint$), along with the Bayesian evidence for each model. We denote these model parameters as ${\bm{\theta}}$ and outline our model on the example of the \ac{TFR}, though the \ac{SN} calibration is near-identical and we derive it in~\cref{sec:supernova_flow}. For a \ac{TFR} catalogue, the observables are the galaxy redshift (converted to the \ac{CMB} frame) $\zobs$, apparent magnitude $\mobs$, and linewidth $\etaobs$. We also introduce two more latent parameters per object: distance $r$ and the ``true'' linewidth $\etatrue$. In the following, we omit the variables that the probabilities do not explicitly depend on to simplify the notation.

The posterior distribution for a single observation (we shall discuss selection later) is then
\begin{equation}\label{eq:posterior}
\begin{split}
    \mathcal{P}(\bm{\theta},\,&r,\,\etatrue \mid \zobs,\,\mobs,\,\etaobs)
    \propto\\
    &\mathcal{L}(\zobs,\,\mobs,\,\etaobs \mid \bm{\theta},\,r,\,\etatrue) \pi(r,\,\etatrue,\,\bm{\theta}),
\end{split}
\end{equation}
where $\mathcal{L}$ is the likelihood of the observables and $\pi$ is the prior on the parameters. Implicitly, the likelihood also depends on the underlying density and velocity field, over which we later marginalise in~\cref{eq:marginalise_field}. We expand the likelihood as
\begin{equation}\label{eq:likelihood}
\begin{split}
    \mathcal{L}(&\zobs,\,\mobs,\,\etaobs \mid \bm{\theta},\,r,\,\etatrue)
    =\\
    &=\mathcal{L}(\zobs\mid \bm{\theta},\,r)
    \mathcal{L}(\mobs\mid \bm{\theta},\,r,\,\etatrue)
    \mathcal{L}(\etaobs\mid \etatrue),
\end{split}
\end{equation}
assuming independence between the observed redshift, apparent magnitude and linewidth. The first term on the right-hand side of~\cref{eq:likelihood} is the likelihood of the observed redshift given the true distance and the flow model. We now introduce the true redshift $\ztrue$ and predicted redshift $\zpred$, where $\zpred$ is a deterministic function of the true distance and the flow model (Eq.~\ref{eq:redshift_addition}). This yields
\begin{equation}
\begin{split}
    \mathcal{L}&(\zobs \mid \bm{\theta},\,r)
    =\\
    &=\iint \mathcal{L}(\zobs \mid \ztrue) p(\ztrue \mid \zpred,\,\bm{\theta})\dd \ztrue \dd \zpred,
\end{split}
\end{equation}
such that
\begin{equation}
    \mathcal{L}(\zobs \mid \ztrue) = \mathcal{N}(c\zobs; c\ztrue, \sigma_{c\zobs}),
\end{equation}
where $\sigma_{c\zobs}$ is the measurement uncertainty of the observed redshift, and
\begin{equation}
    p(\ztrue \mid \zpred,\,\bm{\theta}) = \mathcal{N}(c\ztrue; c\zpred, \sigma_v).
\end{equation}
The true redshift is not equal to the predicted redshift if, for example, the velocity field has some finite resolution. The integral over $\ztrue$ is done analytically and $\zpred$ is a deterministic function of $r$ and the flow model, thus collapsing the integral to $\zpred = \zpred (r,\,\bm{\theta})$ such that
\begin{equation}
    \mathcal{L}(\zobs \mid \bm{\theta},\, r)
    =
    \mathcal{N}\!\left(c\zobs;\, c\zpred,\, \sqrt{\sigma_v^2 + \sigma_{c\zobs}^2}\right).
\end{equation}
The second term in the integrand of~\cref{eq:likelihood} is the likelihood of the observed apparent magnitude given $\etatrue$ and $r$. Through the \ac{TFR}, the absolute magnitude $M$ can be expressed as a function of $\etatrue$ (see~Eq.~\ref{eq:TFR_absmag}), so that the predicted apparent magnitude is $\mpred = M(\etatrue) + \mu(r)$, with $\mu(r)$ being the distance modulus at distance $r$. The corresponding likelihood is
\begin{equation}
    \mathcal{L}(\mobs \mid \bm{\theta},\, r,\, \etatrue)
    =
    \mathcal{N}\!\left(\mobs;\, \mpred,\, \sqrt{\sint^2 + \sigma_m^2}\right),
\end{equation}
where $\sint$ denotes the intrinsic scatter in apparent magnitude (induced by the \ac{TFR}), combined in quadrature with the magnitude measurement error $\sigma_m$, following the same reasoning as for redshift. Finally, the third term is the likelihood of the observed linewidth, which, assuming Gaussian uncertainty on it is simply
\begin{equation}\label{eq:linewidth_likelihood}
    \mathcal{L}(\etaobs\mid \etatrue)
    =
    \frac{\mathcal{N}(\etaobs;\,\etatrue,\,\sigma_{\eta})}{p(S = 1 \mid \hat{\eta},\,w_\eta)},
\end{equation}
where $\sigma_{\eta}$ is the linewidth measurement error and $p(S = 1 \mid \hat{\eta},\,w_\eta)$ accounts for any truncation in $\etaobs$ and is derived (and justified) below in~\cref{eq:eta_truncation}.

Next we consider the prior, which factorises as
\begin{equation}\label{eq:prior_expanded}
    \pi(r,\, \etatrue,\,\bm{\theta})
    = \pi(r \mid \bm{\theta})\, \pi(\etatrue \mid \bm{\theta})\, \pi(\bm{\theta}).
\end{equation}
The first term is the conditional prior on $r$, which account for both the homogeneous Malmquist bias (arising from galaxies being uniformly distributed in volume, so that larger distances are favoured through the $r^2$ volume element) and the inhomogeneous Malmquist bias (arising from large-scale structure). We model it as
\begin{equation}\label{eq:distance_prior}
    \pi(r \mid \bm{\theta})
    = \frac{n(r,\, \bm{\theta})\, f(r,\, \bm{\theta})}
           {\int \mathrm{d}r'\, n(r',\, \bm{\theta})\, f(r',\, \bm{\theta})},
\end{equation}
where $n(r,\, \bm{\theta})$ denotes the number density, $f(r,\, \bm{\theta})$ is defined following~\cite{Virbius} as
\begin{equation}
    f(r,\, \bm{\theta}) = r^p \exp\!\left[-\left(\frac{r}{R}\right)^q\right],
\end{equation}
with $p$, $q$, and $R$ being free model parameters. In particular, it is expected that $p \approx 2$ to reproduce the homogeneous Malmquist bias, the exponential cut-off provides an effective description of selection effects, $R$ sets the scale at which incompleteness becomes significant, and $q$ controls the sharpness of this transition. We emphasise that this is a purely phenomenological approach to modelling selection. A rigorous treatment would require explicitly forward-modelling the survey selection, as outlined by~\cite{Kelly_2008}, but this in turn requires a detailed knowledge of the selection function. Since the \ac{TFR} samples have a complex selection determined by a combination of H\textsc{I} flux and optical magnitude cuts, we instead adopt this phenomenological model. In practice, this distance prior effectively accounts for distance-related selection effects such as cuts in observed redshift or apparent magnitude.

The choice of source number density modelling depends on the reconstruction method. For fields based on linear theory (\citetalias{Carrick_2015}, \citetalias{Courtois2023}), we adopt a linear model,
\begin{equation}
    n(r,\,b_1) = 1 + b_1 \delta(r),
\end{equation}
where $b_1$ is a free parameter and $\delta(r)$ is the density contrast at distance $r$ along the \ac{LOS}. The direct distance tracer samples have different characteristics, which is why we treat $b_1$ as a free parameter. Similarly for~\citetalias{Courtois2023} we sample $b_1$ under a uniform prior. For fields derived from $N$-body simulations (\CBa, \CBb, \citetalias{Sorce_2018}, and~\citetalias{Lilow2024}), we instead use a double-power law model
\begin{equation}
    n(r,\,\rho_t,\,\alpha_{\rm low},\,\alpha_{\rm high}) = \left(\frac{\rho}{\rho_t}\right)^{\alpha_{\rm low}}
           \left(1 + \frac{\rho}{\rho_t}\right)^{\alpha_{\rm high} - \alpha_{\rm low}},
\end{equation}
where $\rho \equiv \rho(r)$ is the density at distance $r$ along the \ac{LOS} to the host, $\alpha_{\rm low}$ and $\alpha_{\rm high}$ are the two slopes, and $\rho_t$ is the transition density, all of which we infer jointly.

Next, in~\cref{eq:prior_expanded} we specify the hyperprior on $\etatrue$, for which we adopt a Gaussian distribution,
\begin{equation}
    \pi(\etatrue \mid \bm{\theta})
    =
    \mathcal{N}\!\left(\etatrue;\, \hat{\eta},\, w_\eta\right),
\end{equation}
where $\hat{\eta}$ and $w_\eta$ denote its mean and width, respectively, and are jointly inferred. Assigning a Gaussian hyperprior to the true (latent) parameter, distinct from the observed quantity, follows the Marginalised Normal Regression framework~\citep{Bartlett_2023}, which has been shown to yield unbiased regression, and is consistent with approaches used in \ac{SN} cosmology~\citep{March_2011, March_2014, Rubin_2015, March_2018, Rubin_2023}. For the remaining parameters, we adopt a reference (scale-invariant) prior $\pi(\sigma_v) \propto 1 / \sigma_v$, applied also to $\sint$, uniform priors on the \ac{TFR} parameters ($\aTFR$, $\bTFR$, $\cTFR$), a reference prior on $w_\eta$, and a uniform prior on $\hat{\eta}$. The external velocity $\bm{V}_{\rm ext}$ is taken to be uniform in both magnitude and direction.

If the observations are independent, the combined likelihood is given by the product over all events. We further account for truncation in the observed linewidth. Following~\cite{Kelly_2008}, selection modifies the posterior for $n$ observed events as
\begin{equation}
    \mathcal{P}(\bm{\theta} \mid \{\bm{d}\})
    \propto
    \frac{\pi(\bm{\theta}) \prod_{i=1}^n \mathcal{L}(\bm{d}_i \mid \bm{\theta})}
         {\left[p(S=1 \mid \bm{\theta})\right]^{-n}},
\end{equation}
where $\pi(\bm{\theta})$ denotes the prior on the model parameters, $\bm{d}_i$ are the data for the $i$\textsuperscript{th} event, and $\mathcal{L}(\bm{d}_i \mid \bm{\theta})$ is the likelihood of $\bm{d}_i$ given parameters $\bm{\theta}$. The factor $p(S=1 \mid \bm{\theta})$ is the fraction of sources that are observed given some total population, and accounts for such ``missing'' data. In the case of the \ac{TFR} model above and truncation in $\etaobs$, this probability is given by
\begin{equation}\label{eq:eta_truncation}
\begin{split}
    p(S = 1 \mid \hat{\eta},\,w_\eta)
    &=
    \iint \mathrm{d}\eta_{\rm obs}\,\mathrm{d}\eta_{\rm true}\;
    p(S = 1 \mid \eta_{\rm obs}) \\
    &\quad\times \mathcal{L}(\eta_{\rm obs} \mid \eta_{\rm true})\,
    \pi(\eta_{\rm true} \mid \hat{\eta},\,w_\eta),
\end{split}
\end{equation}
where $p(S = 1 \mid \eta_{\rm obs})$ is a binary truncation indicator between $\eta_{\min}$ and $\eta_{\max}$,
\begin{equation}
    p(S = 1 \mid \eta_{\rm obs})
    =
    \begin{cases}
        1 & \text{if}\quad \eta_{\min} < \eta_{\rm obs} < \eta_{\max}, \\[6pt]
        0 & \text{otherwise}.
    \end{cases}
\end{equation}
$\mathcal{L}(\eta_{\rm obs} \mid \eta_{\rm true})$ denotes the Gaussian likelihood of the observed given the true linewidth, and $\pi(\eta_{\rm true} \mid \hat{\eta},\,w_\eta)$ is the Gaussian prior on the true linewidth. Given these assumptions, it can be shown that
\begin{equation}
    p(S = 1 \mid \eta_{\rm obs}) = \Phi\!\left(\frac{\eta_{\max} - \eta_{\rm mean}}
                      {\sqrt{\sigma_\eta^{2} + w_\eta^{2}}}\right)
    -
    \Phi\!\left(\frac{\eta_{\min} - \eta_{\rm mean}}
                      {\sqrt{\sigma_\eta^{2} + w_\eta^{2}}}\right),
\end{equation}
where $\Phi(x)$ is the cumulative density function of the standard normal distribution, defined as
\begin{equation}\label{eq:CDF_standard_normal}
    \Phi(x) = \frac{1}{\sqrt{2\pi}} \int_{-\infty}^x e^{-t^2/2} \dd t.
\end{equation}
For this reason, we already included this $p(S = 1 \mid \hat{\eta},\,w_\eta)$ term in~\cref{eq:linewidth_likelihood}. In principle, the same approach should be applied to model selection in optical and H\textsc{I} flux. For instance, the latter would require inferring the H\textsc{I} flux from the H\textsc{I} galaxy mass. However, to avoid this complexity, we instead follow the approach outlined above: selection effects in optical magnitude or H\textsc{I} flux are effectively absorbed into the phenomenological distance prior, while truncation in the linewidth distribution is treated separately. This does not constitute double-counting, as the distance-prior parameterisation is inferred and would therefore favour no cutoff if all selection effects are already accounted for. Furthermore, linewidth selection has a negligible impact on the results presented here.

Bringing this all together produces the posterior:
\begin{equation}\label{eq:full_likelihood}
\begin{split}
    \mathcal{P}(\bm{\theta}\mid &\zobs,\,\mobs,\,\etaobs)
    \propto\\
    &\propto\iint \dd r \dd \etatrue 
    \;\mathcal{N}\;\left(c\zobs; c\zpred, \sqrt{\sigma_v^2 + \sigma_{c\zobs}^2}\right)\\
    &\times\mathcal{N}\left(\mobs; \mpred, \sqrt{\sint^2 + \sigma_{m}^2}\right) \\
    &\times \frac{\mathcal{N}\left(\etaobs; \etatrue, \sigma_{\eta}\right)}{p(S = 1\mid \hat{\eta},\,w_\eta)}\\
    &\times \frac{n(r,\, \bm{\theta})\, f(r,\, \bm{\theta})}{\int \mathrm{d}r'\, n(r',\, \bm{\theta})\, f(r',\, \bm{\theta})} \mathcal{N}\!\left(\etatrue;\, \hat{\eta},\, w_\eta\right) \pi(\bm{\theta}).
\end{split}
\end{equation}
We numerically marginalise over both $r$ and $\etatrue$ for each galaxy at every \ac{MCMC} step. For the radial distance grid, we adopt uniform spacing in comoving distance with step size $0.5~\Mpch$, converting to cosmological redshift or distance modulus using \texttt{Astropy}\footnote{\url{https://www.astropy.org}}~\citep{Astropy1, Astropy2, Astropy3}. For the $\etatrue$ grid, we use 51 uniformly spaced points spanning a range centred on $\etaobs$, with the width defined as ten times $\sigma_\eta$. We explicitly marginalise over both $r$ and $\etatrue$, rather than sampling them with \ac{HMC}, so that the resulting posterior remains relatively low-dimensional and amenable to evidence calculation.

When considering reconstructions with more than one possible realisation of the density and velocity field, we average the posterior to marginalise over the reconstruction as
\begin{equation}\label{eq:marginalise_field}
    \begin{split}
        \mathcal{P}&(\bm{\theta} \mid \zobs,\, \mobs,\, \etaobs) 
        =\\
        &=\int \dd \bm{v}\, \dd \rho \;
        \mathcal{P}(\bm{\theta} \mid \zobs,\, \mobs,\, \etaobs,\, \bm{v},\, \rho)\,
        p(\bm{v}, \rho) \\
        &\approx \frac{1}{N}\sum_{i=1}^{N} 
        \mathcal{P}(\bm{\theta} \mid \zobs,\, \mobs,\, \etaobs,\, \bm{v}_i,\, \rho_i).
    \end{split}
\end{equation}
where $\rho_i$ and $\bm{v}_i$ represent the $i$\textsuperscript{th} density and velocity field realisation, respectively, and $p(\bm{v},\,\rho)$ is the probability of a given realisation (i.e. the posterior distribution from e.g. \texttt{BORG}). We approximate this as a discrete sum over the available realisations, giving each one the same probability. In the case of a \ac{SN} calibration we treat the three independent variables ($m,\,x_1,\,c$) analogously to the \ac{TFR} observables, except we marginalize numerically only over $r$ and explicitly sample the true $x_1$ and $c$ to avoid computing a 3D numerical integral.

To sample the posterior we use the No U-Turns Sampler (NUTS;~\citealt{Hoffman_2011}) method of \ac{HMC}, as implemented in the \texttt{NumPyro} package\footnote{\url{https://num.pyro.ai/en/latest/}}~\citep{Phan_2019}. We remove burn-in and use sufficient steps for the Gelman-Rubin statistic to be one to within $10^{-3}$~\citep{Gelman_1992}. All model parameters, along with their priors, are summarised in~\cref{tab:priors}.

\begin{table*}
    \centering
    \begin{tabularx}{\textwidth}{l X X}
        \toprule
        \textbf{Parameter} & \textbf{Description} & \textbf{Prior} \\[1mm]
        \multicolumn{3}{l}{\textit{Galaxy bias and flow model parameters}} \\ \hline \hline
        $\bm{V}_{\rm ext}$  & External velocity vector & Uniform in both magnitude and direction \\
        $\sigma_v$          & Scatter between the observed and predicted redshifts & $\pi(\sigma_v) \propto 1 / \sigma_v$ \\
        $\beta$ or $\beta^\star$             & Velocity field scaling parameter & Uniform \\
        $b_1$ & Linear galaxy bias parameter (\citetalias{Carrick_2015}, \citetalias{Courtois2023}) & $\pi(b_1) = \mathcal{N}(\beta^\star / f,\,0.04)$ for \citetalias{Carrick_2015}, uniform for~\citetalias{Courtois2023}\\
        $\alpha_{\rm low}$, $\alpha_{\rm high}$, $\ln \rho_t$ & Double power-law galaxy bias parameters (\CBa, \CBb, \citetalias{Sorce_2018}, \citetalias{Lilow2024}) & $\pi(\alpha_{\rm low}) = \mathcal{N}(1, 1)$, $\pi(\alpha_{\rm high}) = \mathcal{N}(0.5, 1)$ (both truncated below at zero) and $\pi(\ln\rho_t) = \mathcal{N}(0, 2)$\\[3mm]
        \multicolumn{3}{l}{\textit{Distance calibration parameters}} \\ \hline \hline
        $\aTFR,\,\bTFR,\,\cTFR$ & TFR coefficients (zero-point, slope, and curvature) & Uniform \\
        $M_{\rm SN},\,\mathcal{A},\,\mathcal{B}$ & Supernova coefficients (absolute magnitude, stretch, and colour) & Uniform \\
        $\sint$ & Magnitude intrinsic scatter & $\pi(\sint) \propto 1 / \sint$ \\
        $R$, $q$, $p$ & Phenomenological distance prior & $\pi(R) = \mathcal{U}(0,100)$ Mpc, $\pi(p) = \mathcal{N}(2,0.1)$ and $\pi(q) = \mathcal{N}(1,30)$\\[3mm]
        \multicolumn{3}{l}{\textit{Latent parameter hyperpriors}} \\ \hline \hline
        $\etatrue$ & True TFR linewidth & $\mathcal{N}(\etatrue \mid \hat{\eta}, w_\eta)$ \\
        $\hat{\eta}$ & Mean of the Gaussian hyperprior on true linewidth & Uniform \\
        $w_\eta$ & Width of the Gaussian hyperprior on true linewidth & $\pi(w_\eta) \propto 1/w_\eta$ \\
        $x_{1, \rm true},\,c_{\rm true}$ & True \ac{SN} stretch and colour parameters & Gaussian hyperpriors with free mean and width \\
        $\bm{\mu}_{\rm SN}$ & Mean of the Gaussian hyperprior on the \ac{SN} stretch, and colour parameters & Uniform \\
        $\mathbf{C}_{\rm SN}$ & Covariance matrix of the Gaussian hyperprior on the \ac{SN} stretch, and colour parameters & $\pi(\sigma_i) \propto 1 / \sigma_i$ for standard deviations $\sigma_i$ and uniform in the correlation coefficient \\ 
        \bottomrule
    \end{tabularx}
    \caption{Summary of the free parameters of our model and their priors.}
    \label{tab:priors}
\end{table*}


\subsection{Evidence calculation}\label{sec:evidence_calculation}

The model evidence $\mathcal{Z}$ is defined as an integral of the product of the likelihood and prior over the parameter space
\begin{equation}\label{eq:evidence_definition}
    \mathcal{Z} = \int \dd \bm{\bm{\theta}} \mathcal{L}(D \mid \bm{\bm{\theta}}) \pi(\bm{\theta}),
\end{equation}
where $D$ is some data and $\bm{\theta}$ are the model parameters. Thus, the interpretation of $\mathcal{Z}$ is that of a likelihood averaged over the prior. The ratio of the evidences for two models, known as the \textit{Bayes factor}, quantifies the relative support of the data for one model over another and serves as the principal model-selection tool of Bayesian statistics, assuming the two models are equally likely \textit{a priori}.

The \ac{TFR} model introduced in~\cref{sec:probabilistic_model} includes, in addition to the model parameters, two (three) latent variables per galaxy in the case of a \ac{TFR} (\ac{SN}) catalogue. This increases the dimensionality of the parameter space to $\mathcal{O}(1000)$, making evidence calculation infeasible using standard methods such as nested sampling~\citep{Skilling_2006}. Recently, machine learning-based methods for evidence calculation, such as \texttt{harmonic}~\citep{McEwan_2021,Polanska_2024} and \textit{floZ}~\citep{Srinivasan_2024}, have been developed. While these methods have shown success in $\mathcal{O}(100)$-dimensional parameter spaces, they have not yet been shown capable of handling the dimensionality that would be required for a full calculation here.

To circumvent this problem, in the \ac{TFR} model we explicitly marginalise over both latent parameters ($r$ and $\etatrue$) at each \ac{MCMC} step, thereby removing them from the posterior and yielding a low-dimensional distribution for which the evidence can be estimated. In contrast, the presence of a third latent parameter in the \ac{SN} model makes explicit numerical marginalisation impractical. Thus, for the \ac{SN} model only we consider an alternative model in which we fix the true values of $x_1$ and $c$ to their observed values, effectively imposing a Dirac distribution prior on them. The observational uncertainties are then propagated not by marginalising over the true values but instead through first-order error propagation, summing them in quadrature with $\sint$ (similar to the method of~\citealt{Boruah_2019}) as
\begin{equation}
    \sint \rightarrow \sqrt{\sint^2 + \mathcal{A}^2\sigma_{x_1}^2 + \mathcal{B}^2\sigma_c^2},
\end{equation}
where $\sigma_{x_1}$ and $\sigma_c$ are the measurement uncertainties of the SALT2 \ac{SN} light curve parameters. This approach reduces the dimensionality of the model to under twenty, and incurs only a small systematic error if the intrinsic scatters of the scaling relations dominate over the measurement uncertainties. In~\Cref{sec:supernova_flow} we use the Foundation sample and compare explicitly the posteriors that the two methods generate. The only parameter that is biased as a result of this assumption is $\mathcal{B}$, whereas all others remain practically unchanged. We emphasise however that this \ac{SN} model approximation is only used for calculating evidences; all other results with the LOSS and Foundation samples use the full model where $x_{1,\mathrm{true}}$ and $c_{\rm true}$ are explicitly sampled (described in~\cref{sec:supernova_flow}).

Now that the dimensionality of the parameter space is sufficiently low, we may apply the \texttt{harmonic}\footnote{\url{https://github.com/astro-informatics/harmonic}} package, which applies normalising flows and the harmonic estimator to calculate the model evidences directly from the \ac{HMC} posterior samples~\citep{McEwan_2021,Polanska_2024}. As can be seen from~\cref{eq:evidence_definition}, the evidence depends on the chosen prior distributions (including the widths of uniform priors). However, in most cases we always compare models with the same parameters, in which case the widths of uniform priors are unimportant. The exception is the model of~\citetalias{Carrick_2015}, which necessitates the additional velocity field scaling parameter $\beta^\star$. Expanding the uniform prior on $\beta$ beyond the range permitted by the likelihood reduces the evidence. Given that the width of the posterior on $\beta$ ranges from $\sim$0.5 in the worst case to $0.05$ in the best, tightening the prior to encapsulate only the posterior would improve the evidence by up to $20-200\times$, or $1.3-2.3$ on a base-10 logarithm scale. This is not significant given the differences in evidence between reconstructions.


\section{Results}\label{sec:results}

We use the following methods to assess the relative goodness of the local Universe reconstructions summarised in~\cref{tab:reconstructions}:
\begin{enumerate}
    \item\label{item:evidence} Bayesian evidence comparison between the different reconstructions for each peculiar velocity (Section~\ref{sec:evidence_comparison}).
    \item\label{item:sigmav} The magnitude of the inferred $\sigma_v$ (Section~\ref{sec:sigma_v_consistency}).
    \item\label{item:beta} Difference between the inferred $\beta$ (see Eq.~\ref{eq:predicted_peculiar_velocity}) and unity in cases where the reconstructions are derived from the matter field (Section~\ref{sec:velocity_scaling}).
    \item\label{item:vext} Comparison of the inferred $\bm{V}_{\rm ext}$ and the agreement of the reconstructed velocity at $r\rightarrow0$ with the Local Group velocity in the \ac{CMB} frame (Section~\ref{sec:external_flow} and Section~\ref{sec:bulk_flows}).
\end{enumerate}

The interpretation of these metrics is as follows. The evidence gives the relative probability of the data under the model, which as described above is the main goodness-of-fit statistic of Bayesian inference.~\ref{item:sigmav} describes the magnitude of residuals between the observed and predicted redshifts, which measures the accuracy of the model in the posterior region of parameter space. We will also show it to correlate well with the evidence.~\ref{item:beta} also indicates whether the predicted velocities are in broad agreement with the observations: if $\beta<1$ then the velocities are over-predicted on average while if~$\beta>1$ they are underpredicted, a systematic bias in either case. Lastly,~\ref{item:vext} measures the velocity at which the data requires the reconstruction box to move relative to the \ac{CMB}. A smaller $\bm{V}_{\rm ext}$ is preferred, especially when the reconstruction volume is large, as this implies that the velocities within the box are sufficient to account for the data. However, $\bm{V}_{\rm ext}$ depends on both the reconstruction and the peculiar velocity sample used to constrain it, and thus it can only serve as a qualitative comparison.
The dependence on the reconstruction volume arises partly from the fact that if the velocities within the box are sourced solely by internal matter, their average will be zero when the entire box is enclosed. Consequently, the average enclosed velocity in the \ac{CMB} frame (the bulk flow) at such radii is determined exclusively by $\bm{V}_{\rm ext}$ which we assume to be constant. The bulk flow magnitude follows a Maxwell-Boltzmann distribution in the \ac{LCDM} model (see e.g. Appendix B of~\citealt{Davis_2011}). In~\cref{sec:bulk_flows} we also present and compare the radial dependence of the predicted bulk flows and in~\cref{sec:growth_rate} we report constraints on the growth-rate of large-scale structure using~\citetalias{Carrick_2015} and the \ac{CF4} \ac{TFR} samples.

In addition to using the individual peculiar velocity samples outlined in~\cref{tab:pv_samples}, we also construct a single joint dataset by combining the LOSS, Foundation, and both \ac{CF4} \ac{TFR} samples. For the latter, we prioritise \ac{SDSS} photometry when available. We do not include 2MTF or~\SFI~in this joint catalogue as these were used in constructing \ac{CF4}; all objects in our joint dataset therefore contain just one measurement per galaxy. We select only objects with $\zCMB < 0.05$ that are within the reconstructed volume.


\subsection{Evidence comparison}\label{sec:evidence_comparison}

\begin{figure*}
    \centering
    \includegraphics[width=\textwidth]{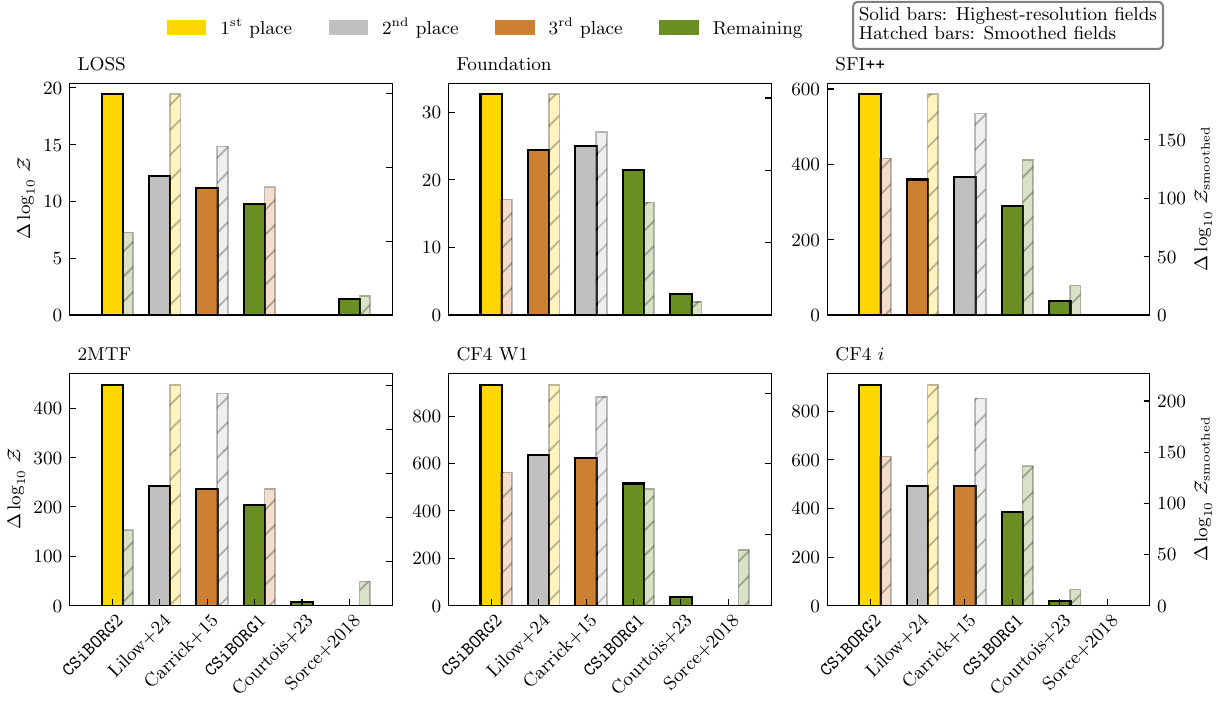}
    \caption{Differences in logarithmic evidences $\mathcal{Z}$ from our flow validation model for various local Universe reconstructions (shown on the $x$-axis, see~\cref{tab:reconstructions}), compared against peculiar velocity samples (individual panels, see~\cref{tab:pv_samples}).~\emph{Higher bars indicate a preferred model,} and a bar of zero height indicates the reference (least successful) model. The logarithmic evidences are normalised with respect to the reference model as only relative differences are meaningful. Solid bars show evidences using the highest available resolution for each model, while hatched bars show evidences when all fields are smoothed to the resolution of $7.8~\Mpch$, twice that of~\protect\cite{Courtois2023}. Overall, \CBb\ is the preferred model while the CosmicFlows-based reconstructions (\protect\citealt{Sorce_2018} and \protect\citealt{Courtois2023}) are disfavoured. Upon smoothing, the reconstruction of~\protect\cite{Lilow2024} becomes marginally preferred.}
    \label{fig:lnZ_comparison}
\end{figure*}

In~\cref{fig:lnZ_comparison}, we show the Bayesian evidences to compare the relative goodness-of-fit of reconstructions against the peculiar velocity samples. Since the absolute value of the evidence of a model is meaningless, we only show the differences of the logarithmic evidences $\Delta \log_{10} \mathcal{Z}$, relative to the model with the lowest evidence. The \ac{TFR} samples have greater discriminating power (larger differences in $\log_{10} \mathcal{Z}$) compared to \ac{SN} samples, since despite their smaller intrinsic scatter, the \ac{SN} samples that we use contain significantly fewer objects. For all models except that of~\citetalias{Carrick_2015} we fix $\beta = 1$ for this calculation. We find \CBb\ to be the best model across all peculiar velocity samples, with~\citetalias{Lilow2024} typically ranking second and~\citetalias{Carrick_2015} third with only a small difference between these latter two. Relative to these, the two reconstructions based on CosmicFlows,~\citetalias{Sorce_2018} and~\citetalias{Courtois2023}, are strongly disfavoured. When testing the joint \ac{SN} and \ac{TFR} dataset, we find qualitatively similar results.

A possible driver of the evidence ratios is the varying resolutions of the reconstruction. To isolate this effect, we consider smoothing all fields to a resolution of $7.8~\Mpch$ which is twice the voxel resolution of~\citetalias{Courtois2023}. This is achieved by applying a Gaussian kernel with a standard deviation of $\sqrt{(7.8~\Mpch)^2- \sigma_0^2}$ to the densities and velocities, where $\sigma_0$ is the original resolution (voxel size) of the field. However, a Gaussian smoothing kernel a standard deviation $\sigma_0$ is not equivalent to a field with voxel resolution $\sigma_0$, thus this ``matching'' of resolutions is only approximate. We then recalculate the evidences for the smoothed fields, showing the results as hatched bars in~\cref{fig:lnZ_comparison}. This reduces the differences in $\log_{10} \mathcal{Z}$ between the reconstructions, and also causes~\citetalias{Lilow2024} to become the preferred model although both \ac{CSIBORG} fields continue to perform well. The preference for~\citetalias{Lilow2024} upon smoothing might arise from accurately capturing large-scale flows, as indicated by the relatively small magnitude of $\bm{V}_{\rm ext}$ discussed in~\cref{sec:external_flow}. Even after smoothing the CosmicFlows-based reconstructions remain strongly disfavoured. This indicates that the finer-grained information in the other reconstructions contains accurate information, and that the redshift-space positions of \TWOMPP\ galaxies afford better constraints than the peculiar velocities of CosmicFlows.

In~\cref{fig:smoothing_effect} we investigate the effect of smoothing on~\citetalias{Carrick_2015} using the joint \acp{SN} and \ac{TFR} dataset. We plot both $\Delta \log_{10} \mathcal{Z}$ and $\sigma_v$ as functions of smoothing scale $R_{\rm smooth}$. As expected, the goodness-of-fit degrades with increasing smoothing scale. Notably, $\sigma_v$ increases proportionally, suggesting it is a good indicator of goodness-of-fit. However, $\sigma_v$ should not be compared between different peculiar velocity samples, as it depends on sample properties such as the relative prevalence of field and cluster galaxies. {We infer from these results that the field of~\citetalias{Carrick_2015} would need to be smoothed with a scale of $R_{\rm smooth} \approx 22~\Mpch$ to match the evidence of~\citetalias{Courtois2023}, since we find $\Delta \log_{10}\mathcal{Z}$ of $890$ between~\citetalias{Carrick_2015} and~\citetalias{Courtois2023}.

\begin{figure}
    \centering
    \includegraphics[width=\columnwidth]{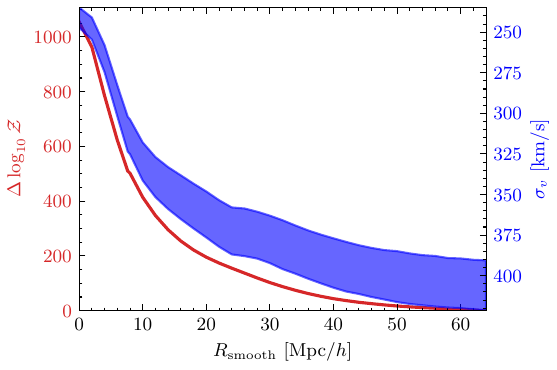}
    \caption{Differences in logarithmic evidence ($\Delta \log_{10} \mathcal{Z}$, with higher values indicating a better goodness-of-fit) and inferred $\sigma_v$ (scatter between observed and predicted redshifts) as functions of smoothing scale $R_{\rm smooth}$ for the joint \ac{SN} and \ac{TFR} dataset against~\protect\cite{Carrick_2015}. The $\sigma_v$ axis is inverted. Both the goodness-of-fit, and $\sigma_v$ quickly degrade with increased smoothing, indicating that $\sigma_v$ is an effective goodness-of-fit indicator.}
    \label{fig:smoothing_effect}
\end{figure}


\subsection{Magnitude of redshift residuals}\label{sec:sigma_v_consistency}

\begin{figure}
    \centering
    \includegraphics[width=\columnwidth]{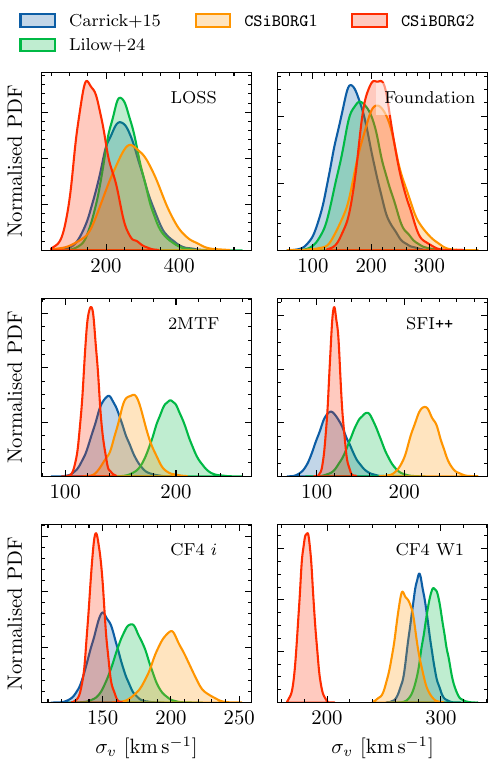}
    \caption{The posteriors on $\sigma_v$ per reconstruction applied to each dataset. As expected, reconstructions favouring lower $\sigma_v$ tend to be preferred in terms of evidence as well (see~\cref{fig:lnZ_comparison}). \CBb\ and~\protect\cite{Carrick_2015} produce the lowest $\sigma_v$ values on the \ac{CF4} datasets. For clarity, we omit $\sigma_v$ for~\protect\cite{Courtois2023} and~\protect\cite{Sorce_2018}, as their values are substantially higher. For example, in both 2MTF and CF4 $i$-band $\sigma_v \approx 500~\kmsec$.}
    \label{fig:sigmav_comparison}
\end{figure}

We now turn to $\sigma_v$, which measures the residuals between the observed and predicted redshifts. This contains information on how well the reconstruction reproduces the ``observed'' velocity field of the local Universe. $\sigma_v$ depends on whether galaxy non-linear velocities were suppressed by grouping. While it is preferable for $\sigma_v$ to be lower, there must be a lower floor (per sample) describing the effect of small-scale motions that the reconstructed fields cannot hope to capture. This floor would be raised by any unaccounted-for systematic effects or other forms of model misspecification.

In~\cref{fig:sigmav_comparison} we present the constraints on $\sigma_v$ using each of the velocity fields. Generally, reconstructions with lower $\sigma_v$ values also achieve higher evidence scores, while those with higher $\sigma_v$ are less favoured. However, the two \ac{SN} samples alone lack sufficient data to provide meaningful constraints on $\sigma_v$. For the~\ac{TFR} samples, it is notable that~\citetalias{Carrick_2015} yields marginally lower $\sigma_v$ values compared to~\citetalias{Lilow2024}, despite with the evidence comparison's weak preference for~\citetalias{Lilow2024}. This, together with the fact that the evidence comparison does not vastly prefer~\citetalias{Lilow2024} over~\citetalias{Carrick_2015}, appears to indicate that the ``augmentation'' to~\citetalias{Carrick_2015} provided by learning is not particularly successful. The \CBb\ velocities clearly produce the highest evidences and often require the lowest $\sigma_v$.

In contrast, reconstructions based on CosmicFlows data require the largest $\sigma_v$, even for the \ac{CF4} samples. Owing to their large values relative to the other models, we do not show them explicitly in~\cref{fig:sigmav_comparison}. However, in the \ac{CF4} W1 sample we find $\sigma_v = 387 \pm 17~\kmsec$ for~\citetalias{Courtois2023} and $395 \pm 15~\kmsec$ for~\citetalias{Sorce_2018}. For the \ac{CF4} $i$-band sample and for 2MTF, $\sigma_v$ is around $500~\kmsec$ in both cases. This outcome is surprising as these reconstructions are directly constrained by CosmicFlows data, of which the \ac{TFR} samples are a subset.


\subsection{Consistency of the velocity scaling}\label{sec:velocity_scaling}

\begin{figure}
    \centering
    \includegraphics[width=\columnwidth]{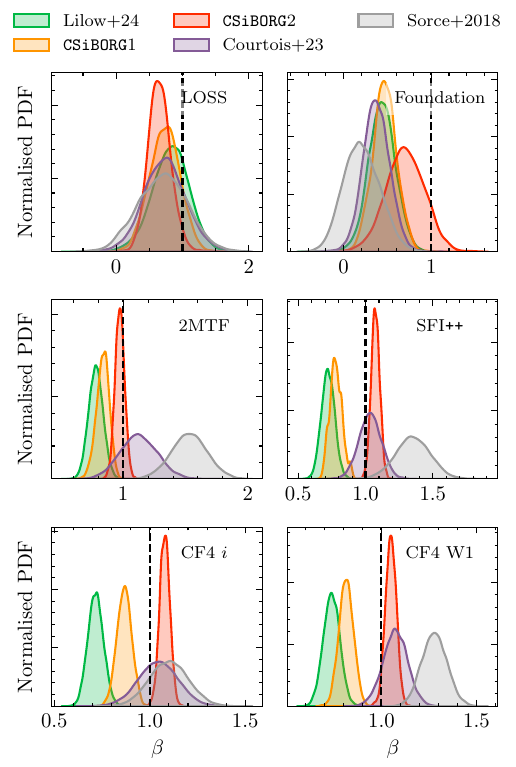}
    \caption{Inferred values of $\beta$ (Eq.~\ref{eq:predicted_peculiar_velocity}) given a uniform prior, for the models in which velocities are derived from the matter field. The fiducial value $\beta=1$ is that at which the velocities prefer not to be rescaled. Both~\protect\cite{Lilow2024} and \CBa\ systematically overpredict velocities across all samples ($\beta < 1$), whereas the evidence-preferred model, \CBb, is consistent with unity in all cases except for the \ac{CF4} $i$-band, where the deviation remains below $3\sigma$.}
    \label{fig:beta_comparison}
\end{figure}

Next we examine $\beta$, which scales the velocity field before comparison to the data (Eq.~\ref{eq:predicted_peculiar_velocity}). Here we treat $\beta$ as a free parameter with a uniform prior, with the expectation that $\beta = 1$ for reconstructions where the velocities are sourced by the matter field; a field is better behaved to the extent that this is preferred \textit{a posteriori}.

The constraints on $\beta$ are shown in~\cref{fig:beta_comparison}. We find that the model preferred by the Bayesian evidence, \CBb, somewhat underpredicts velocities only in the \SFI\ and \ac{CF4} $i$-band samples, where $\beta = 1.13 \pm 0.03$ and $1.14 \pm 0.02$. However, even in these cases the deviations are 2.3 and $3\sigma$, respectively, while all other samples are consistent with $\beta = 1$ to within $1.5\sigma$. (We note that in our previous evidence comparison we had fixed $\beta = 1$.) In contrast, \CBa\ yields $\beta$ values inconsistent with unity at more than $3\sigma$ in all cases except for the LOSS catalogue. Similarly,~\citetalias{Lilow2024} consistently produces $\beta<1$ on all \ac{TFR} samples, suggesting its reconstructed velocities are too high or otherwise systematically biased. On the other hand, \citetalias{Sorce_2018} has $\beta$ substantially higher ($>2.5\sigma$) on all \ac{TFR} samples except \ac{CF4} $i$-band. \citetalias{Courtois2023} is consistent with unity in all cases, except the Foundation sample. We leave a detailed discussion of the velocity scaling in~\citetalias{Carrick_2015} to~\cref{sec:growth_rate}, as in~\citetalias{Carrick_2015} it is directly related to the growth rate of large-scale structure.


\subsection{Magnitude of the external velocity}\label{sec:external_flow}

\begin{figure}
    \centering
    \includegraphics[width=\columnwidth]{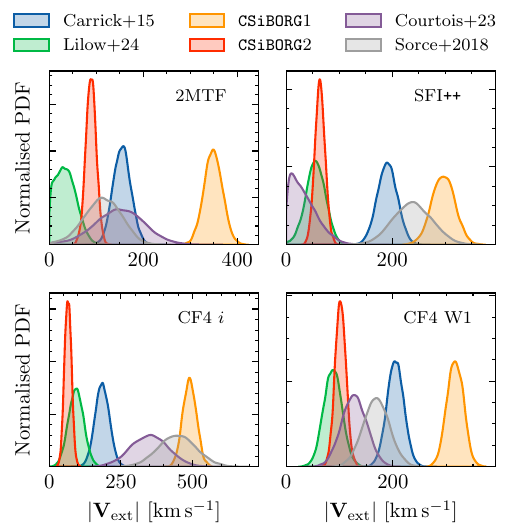}
    \caption{Inferred values of the magnitude of $\bm{V}_{\rm ext}$ (Eq.~\ref{eq:predicted_peculiar_velocity}). A higher magnitude indicates that the reconstruction box is moving faster relative to the rest of the Universe.}
    \label{fig:Vext_comparison}
\end{figure}

Next we study $|\bm{V}_{\rm ext}|$, the speed of the entire reconstruction box in the \ac{CMB} frame sourced by matter outside it. This depends on both the reconstruction method and the peculiar velocity survey, which sets the effective probed volume.

The posteriors on $|\bm{V}_{\rm ext}|$ are shown in~\cref{fig:Vext_comparison}. We observe that although~\citetalias{Carrick_2015} and~\citetalias{Lilow2024} yield similar goodness-of-fit values, the latter requires a significantly smaller external flow despite both reconstructions being constrained by (almost) the same data. The external flow magnitude in~\citetalias{Lilow2024} is nearly consistent with zero for the 2MTF, \SFI, and \ac{CF4} \ac{TFR} $i$-band samples, with values of $35 \pm 21$, $55 \pm 19$, and $93 \pm 26~\kmsec$, respectively, since their reconstruction already includes $\bm{V}_{\rm ext}$-like contribution by construction. In contrast,~\citetalias{Carrick_2015} requires larger magnitudes of $154 \pm 19$, $190 \pm 18$, and $184 \pm 24~\kmsec$ for those samples.

Among the other reconstructions, \CBb\ generally requires the smallest external flow magnitude ($100~\kmsec$), whereas \CBa\ typically requires a larger speed of approximately $400~\kmsec$. For the CosmicFlows-based reconstructions, we find that their preferred flow magnitudes disagree significantly only in the case of \SFI, with~\citetalias{Sorce_2018} requiring a notably higher magnitude. Although in other cases they agree well, the magnitude required by~\citetalias{Sorce_2018} is systematically higher, potentially due to the shallower depth of the CosmicFlows-2 data. However, for both \ac{CF4} samples,~\citetalias{Sorce_2018} and~\citetalias{Courtois2023} require a higher flow magnitude than \CBb (or~\citetalias{Lilow2024}).
 

\subsection{Bulk flow curve and Local Group velocity}\label{sec:bulk_flows}

\begin{figure}
    \centering
    \includegraphics[width=\columnwidth]{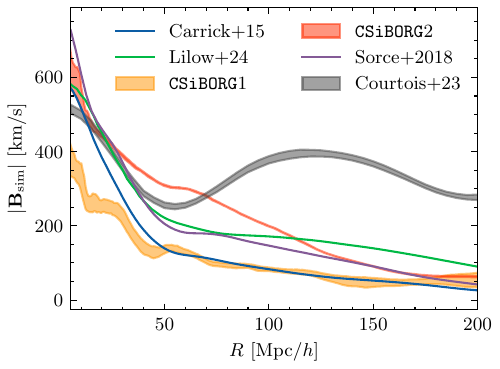}
    \caption{The bulk flow magnitude measured directly in the simulation rest frame (i.e., without considering additional $\bm{V}_{\rm ext}$) as a function of radius $R$. All reconstructions, except for~\protect\cite{Courtois2023}, exhibit a decreasing trend as expected in~\ac{LCDM}. For reconstructions that provide posterior samples, we plot the 16\textsuperscript{th} and 84\textsuperscript{th} percentiles as bands.}
    \label{fig:BF_sim}
\end{figure}

Finally, we calculate the bulk flows produced by the various reconstructions. The velocities in each reconstruction are generated by the internal matter distributions and thus are in the ``simulation'' rest frame, where the average velocity of the box is zero by definition. The exception to this is~\citetalias{Lilow2024} whose neural network predicts both the density and velocity field, without explicitly computing the latter from the former and, thus, can learn ``super-survey'' scales encoded in the redshift-space distortions. This means that the~\citetalias{Lilow2024} velocities already have $\bm{V}_{\rm ext}$ encoded in them, and hence the fiducial value for our free parameter would be zero. We denote the simulation bulk flow as $\bm{B}_{\rm sim}$ and define it according to the standard bulk flow definition~\citep{Watkins_2009,Feldman_2010,Nusser_2011,
Hoffman_2015,Watkins_2015,Nusser_2016,Scrimgeour_2016,Feix_2017,Hellwing_2018,Peery_2018,Whitford_2023,Watkins_2023}
\begin{equation}
    \bm{B}_{\rm sim}(R) = \frac{\beta}{V} \int_V \bm{v}(\bm{r}) \dd V,
\end{equation}
where $V$ is the volume enclosed within a sphere of radius $R$ around the observer at the centre of each box. To calculate this we again set $\beta = 1$ for all reconstructions except~\citetalias{Carrick_2015}, where we use the inferred posterior samples of $\beta$ which we discuss in~\cref{sec:growth_rate}.

In~\cref{fig:BF_sim} we show the simulation-frame $\bm{B}_{\rm sim}(R)$. All reconstructions except for~\citetalias{Courtois2023} display a decreasing trend broadly consistent with the \ac{LCDM} expectation. \citetalias{Carrick_2015} assumes only the continuity equation to relate the galaxy density field to the velocity field, treating the proportionality constant ($\beta^\star$) as a free parameter. By contrast, other reconstructions (partially) assume \ac{LCDM} during the process itself: \citetalias{Lilow2024} uses a \ac{LCDM} training set, \ac{BORG} uses a \ac{LCDM} prior on the \acp{IC} and a \ac{LCDM} gravity model,~\citetalias{Courtois2023} uses a \ac{LCDM} prior on the $z = 0$ density field and linear theory to calculate velocities, and~\citetalias{Sorce_2018} assumes \ac{LCDM} when constraining the \acp{IC} with the reverse Zel'dovich approximation (and then to simulate the reconstruction). Most reconstructions show a bulk flow magnitude approaching approximately $600~\kmsec$ near the origin, except for \CBa, which reaches only $400~\kmsec$.

\begin{figure*}
    \centering
    \includegraphics[width=\textwidth]{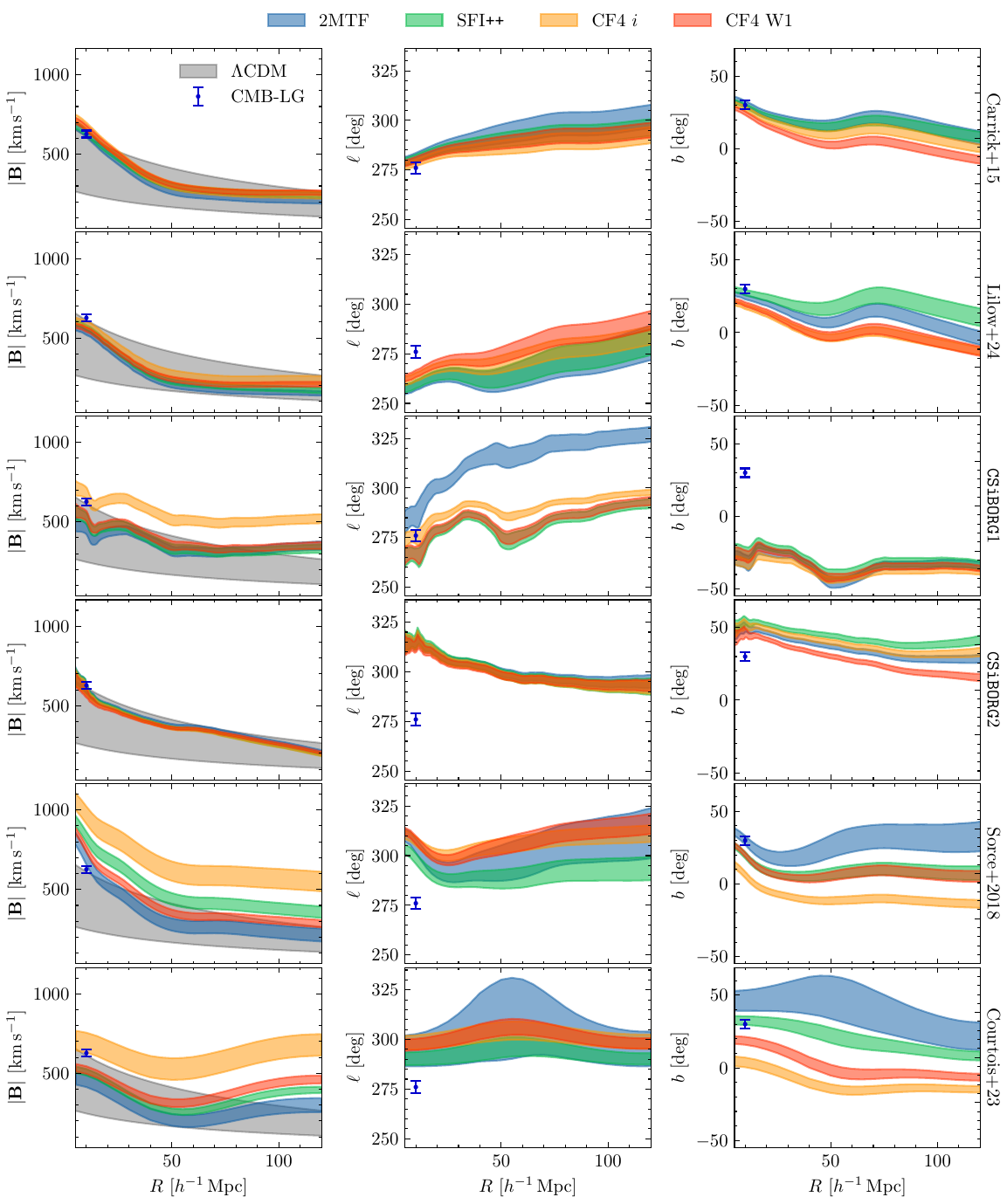}
    \caption{The bulk flow magnitude (left column) and direction in galactic longitude ($\ell$; centre column) and latitude ($b$; right column) as a function of radial distance for each reconstruction (rows). The bulk flow is computed as $\bm{B}_{\rm sim} + \bm{V}_{\rm ext}$, where $\bm{B}_{\rm sim}$ is the bulk flow within each box (Fig.~\ref{fig:BF_sim}), and $\bm{V}_{\rm ext}$ is the external velocity constrained by the displayed peculiar velocity samples. The grey band represents the 16\textsuperscript{th} to 84\textsuperscript{th} percentile range for the bulk flow magnitude expected in \ac{LCDM}, while the single dark blue $1\sigma$ error bar indicates the motion of the Local Group in the \ac{CMB} frame~\citep{Planck_2020}. This is plotted slightly offset from $R = 0$ simply for visual clarity.}
    \label{fig:BF_CMB}
\end{figure*}

Next we study the radial dependence of the bulk flow in the \ac{CMB} frame, which we calculate as
\begin{equation}
    \bm{B}(R) = \bm{B}_{\rm sim}(R) + \bm{V}_{\rm ext}.
\end{equation}
Note that this exhibits radial variation only within the reconstruction box: beyond it $\bm{B}(R)$ equals the constant $\bm{V}_{\rm ext}$. In~\cref{fig:BF_CMB} we present the \ac{CMB} frame bulk flow magnitude and direction in galactic coordinates. We overplot the velocity of the Local Group in the \ac{CMB} frame (on which more below) as well as the $1\sigma$ expectation for the bulk flow magnitude in \ac{LCDM} using a top-hat window function of radius $R$ (see e.g. Sec.~6.1.2 of~\citealt{Boruah_2019}). We find that the bulk flow magnitude for most reconstructions aligns well with the \ac{LCDM} expectation. This is not true for \CBa\ and~\citetalias{Courtois2023}, however, which overpredict the bulk flow at large $R$. In \CBa this is driven by the large $|\bm{V}_{\rm ext}|$ (Fig.~\ref{fig:Vext_comparison}), while for~\citetalias{Courtois2023} it is driven by the large internal bulk flow (\cref{fig:BF_sim}). Although this trend does not indicate significant tension with \ac{LCDM} at the radii we probe ($\lesssim 150~\Mpch$), Fig.~3 of~\cite{Courtois2023} shows that their bulk flow appears to continue increasing beyond the radii we plot, posing a greater challenge to \ac{LCDM} if true. This is also evident in~\citet{Watkins_2023}, based on the same data.

Examining the bulk flow direction as a function of radial distance, we observe that the reconstructions of~\citetalias{Carrick_2015} and \citetalias{Lilow2024} display similar trends, though they do differ in longitude. The two \ac{CSIBORG} reconstructions show significant disagreement in latitude. This disagreement between the two \ac{BORG} reconstructions is not entirely surprising. The \ac{IC} inference of \CBa used an inferior gravity model and a different galaxy bias model~\citep{Stopyra_2023}, leading to an overestimation of cluster masses, which in turn should affect the local velocity field. Specifically, \CBa\ used a 10-step particle-mesh solver for the gravitational dynamics, whereas \CBb\ adopted a 20-step \texttt{COLA} integrator. The impact of gravity model accuracy on the \ac{BORG} \acp{IC} is discussed in detail by~\citet{Stopyra_2023}. This is reflected in the evidence comparison, where \CBa\ is strongly disfavoured relative to \CBb. Finally, the CosmicFlows-based reconstructions demonstrate similar directional trends, but the preferred $\bm{V}_{\rm ext}$ varies significantly between the peculiar velocity samples leading to more prominent differences in bulk flow direction across samples.

Finally we investigate consistency with the one piece of information we have about the bulk flow curve: that the velocity of the \ac{LG} in the \ac{CMB} frame is $620 \pm 15~\kmsec$ towards $(\ell, b) = (271.9^\circ \pm 2.0^\circ, 29.6^\circ \pm 1.4^\circ)$, as inferred from the \ac{CMB} dipole~\citep{Planck_2020}. To recover the predicted \ac{LG} velocity $\bm{B}_{\rm LG}$, we evaluate the velocity fields at the origin to obtain $\bm{B}_{\rm sim}(R=0)$ and add to it $\bm{V}_{\rm ext}$. However, this cannot be directly compared to the observed \ac{CMB}--\ac{LG} velocity due to the $\sigma_v$ scatter describing a point-by-point mismatch between predicted and observed velocities. To account for this we calculate $\bm{B}_{\rm LG}$ as
\begin{equation}\label{eq:BF_LG}
    \bm{B}_{\rm LG} = \bm{B}_{\rm sim}(R=0) + \bm{V}_{\rm ext} + \bm{V}_{\rm rand},
\end{equation}
where $\bm{V}_{\rm rand}$ is a random vector with components drawn from a mean-zero Gaussian with standard deviation $\sigma_v$ in each component. The result is shown in~\cref{fig:BF_CMB_origin}.

\begin{figure*}
    \centering
    \includegraphics[width=\textwidth]{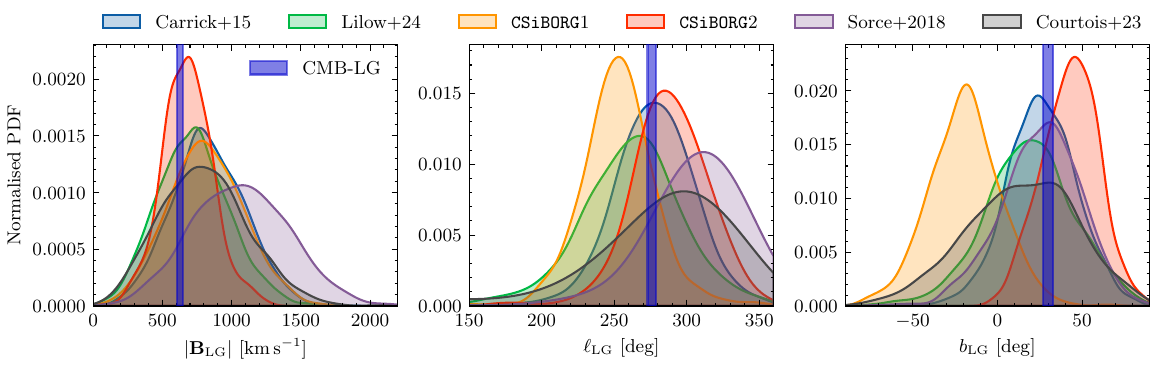}
    \caption{The predicted \ac{LG} velocity following~\cref{eq:BF_LG}, where we propagate the $\sigma_v$ noise. We show its magnitude, longitude, and latitude ($|\bm{B}_{\rm LG}|,\,\ell_{\rm LG},\,b_{\rm LG}$) in the \ac{CMB} frame in galactic coordinates calibrated against the \ac{CF4} \ac{TFR} W1-band sample for the various reconstructions. All predicted \ac{LG} velocities are consistent with the observed \ac{CMB}-\ac{LG} velocity ($1\sigma$ region in dark blue rectangles,~\protect\citealt{Planck_2020}), although \CBa shows a $\sim$2$\sigma$ deviation in latitude.}
    \label{fig:BF_CMB_origin}
\end{figure*}

We find that propagating the $\sigma_v$ uncertainty significantly broadens the uncertainties on the predicted \ac{LG} motion compared to when this scatter is not included, since its value is comparable to the \ac{LG}--\ac{CMB} speed. Additionally, the variations between the different peculiar velocity samples, each with its specific $\bm{V}_{\rm ext}$ and $\sigma_v$, are no longer significant. Therefore, in~\cref{fig:BF_CMB_origin} we present the predicted \ac{LG} motion for the \ac{CF4} \ac{TFR} W1 sample only, which is representative of all samples' results. We find agreement with the observed \ac{CMB}-\ac{LG} velocity in both magnitude and direction for all models; the only minor tension is found in \CBa, where the predicted latitude deviates from the \ac{CMB} dipole by $\sim$2$\sigma$ ($-22^\circ \pm 21^\circ$).

\subsection{Growth rate measurement}\label{sec:growth_rate}

In the reconstruction of~\citetalias{Carrick_2015}, the posterior samples on $\beta^\star$, the velocity scaling parameter, are related to the parameter combination $f\sigma_{8, {\rm NL}}$ through~\cref{eq:beta_star}. As many cosmological probes measure the amplitude of the linear field $\sigma_{8,\mathrm{L}}$, we wish to convert this to $f\sigma_{8,\mathrm{L}}$. A mapping between $\sigma_{8,\mathrm{NL}}$ and $\sigma_{8,\mathrm{L}}$ was derived in~\cite{Juszkiewicz_2010}, but we derive an alternative mapping using the non-linear matter power spectrum. To this end, we first obtain $\sigma_{8, {\rm NL}}$ from $f\sigma_{8, {\rm NL}}$ by using the approximation $f \approx \Omega_{\rm m}^{0.55}$~\citep{Wang_1998}. For all these conversions, we assume a flat \ac{LCDM} with the cosmological parameters $h = 0.6766$, $\Om=0.3111$, $\Omega_{\rm b}=0.02242 / h^2$ and $n_s=0.9665$, except the amplitude of the primordial power spectrum $A_{\rm s}$ which we wish to infer. This is achieved by running a root-finding algorithm, where, for a given $A_{\rm s}$, $\sigma_{8, {\rm NL}}$ is computed by performing the appropriate integral of the non-linear matter power spectrum, which is evaluated using the \texttt{syren-new} emulator\footnote{\url{https://github.com/DeaglanBartlett/symbolic_pofk}}~\citep{Bartlett_2024,Sui_2024}. Once the corresponding value of $A_{\rm s}$ has been obtained, this is converted to the linear value of $\sigma_{8,\mathrm{L}}$ using the conversion given in Eq.~5 of~\citet{Sui_2024}. For our chosen cosmological parameters, we find a correction to the mapping previously derived by~\cite{Juszkiewicz_2010}: the non-linear matter power spectrum-based mapping from $\sigma_{8,\mathrm{NL}}$ to $\sigma_{8,\mathrm{L}}$ is approximately three per cent lower than reported by~\cite{Juszkiewicz_2010}. The exact value of the discrepancy ranges from approximately 3-5 per cent, depending on the exact non-linear prescription used. For example, using the~\citet{Mead_2016} implementation of \texttt{hmcode} gives a value which is five per cent lower than the~\citeauthor{Juszkiewicz_2010} formula. We choose to report values using the \texttt{syren-new} approximation of the non-linear power spectrum for this conversion due to its better agreement with $N$-body simulations~\citep{Bartlett_2024,Sui_2024} compared to \texttt{hmcode} and \texttt{halofit} \citep{Smith_2003,Takahashi_2012}. We show this mapping in~\cref{fig:sigma8_mapping}. Similarly,~\cite{Hollinger_2024} derived the non-linear to linear $\sigma_8$ mapping using~\texttt{halofit}~\citep{Smith_2003,Takahashi_2012} and reported comparable disagreements with the results of~\cite{Juszkiewicz_2010}.

\begin{figure}
    \centering
    \includegraphics[width=\columnwidth]{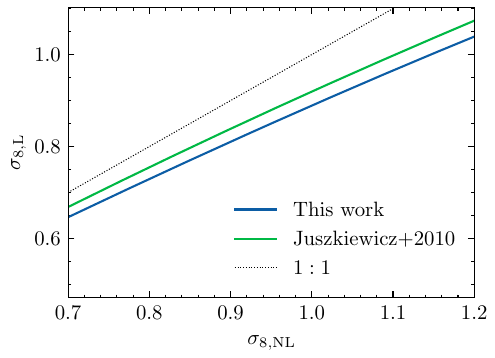}
    \caption{The mapping from $\sigma_{8, {\rm NL}}$, the amplitude of density fluctuations computed using the non-linear matter power spectrum compared to the linearly evolved $\sigma_{8, {\rm L}}$. For our choice of cosmological parameters ($h = 0.6766$, $\Om=0.3111$, $\Omega_{\rm b}=0.02242 / h^2$ and $n_s=0.9665$), we find that the resulting $\sigma_{8,\mathrm{L}}$ is $\sim$3 per cent lower than the one derived by~\protect\cite{Juszkiewicz_2010}.}
    \label{fig:sigma8_mapping}
\end{figure}

In~\cref{fig:C15_beta}, we compare our measurements of $\beta^\star$ to various previous calibrations of~\citetalias{Carrick_2015}. We find $\beta^\star = 0.450 \pm 0.013$ for the joint sample (LOSS, Foundation+\ac{CF4} \ac{TFR} $i$-band, \ac{CF4} \ac{TFR} W1-band). This measurement is under the assumption of a linear bias parameter $b_1$, which is inferred jointly with an uninformative prior. By contrast, assuming a power-law bias in density would systematically underpredict $\beta^\star$, since the model would inflate the probability of galaxies residing in overdense regions, where peculiar velocities are naturally higher, and thus attempt to lower $\beta^\star$. As~\citetalias{Carrick_2015} is a linear-theory reconstruction with an effective smoothing scale of $4~\Mpch$, a linear bias is more consistent with the assumptions underlying it.

The measured $\beta^\star$ is similar to both the W1 and $i$-band-only measurement individually. We summarise the inferred values of $\beta^\star$ in~\cref{tab:C15_beta}. The joint sample includes approximately twice as many data points, which, as expected, reduces the uncertainty in $\beta^\star$ by roughly a factor of $\sqrt{2}$. As shown in~\cref{fig:C15_beta}, our measurement is only marginally higher than those reported by~\cite{Carrick_2015} and~\cite{Boruah_2019}, but significantly lower than the values of~\citet{Said_2020} and~\citet{Boubel_2024}, suggesting inconsistency at the level of $>4\sigma$. The difference is particularly noteworthy for~\cite{Boubel_2024}, who also used the \ac{CF4} \ac{TFR} sample, whereas~\cite{Said_2020} used fundamental plane data. One key difference is that~\cite{Said_2020}, in their Eq.~23, adopt a distance prior that effectively assumes a galaxy bias of $b_1 = 1$, i.e. that the peculiar velocity sample shares the same bias as the \TWOMPP\ sample, whereas we consistently infer $b_1 > 1$ (see~\cref{tab:C15_beta}). It is noteworthy that the \SFI\ sample alone favours relatively small values of $\beta^\star$, consistent with~\cite{Said_2020} and~\cite{Boubel_2024}. However, because a subset of \SFI\ sources is included in the \ac{CF4} compilation, we exclude it from the joint constraint. This discrepancy may indicate a systematic effect in the \SFI\ data.

\begin{table}
\centering
\begin{tabular}{lccc}
Catalogue & $\beta^\star$ & $S_8$ & $b_1$ \\
\hline
2MTF           & 0.455 $\pm$ 0.023 & 0.787 $\pm$ 0.046 & 1.208 $\pm$ 0.030 \\
SFI\texttt{++} & 0.369 $\pm$ 0.020 & 0.650 $\pm$ 0.041 & 1.212 $\pm$ 0.018 \\ \hline \hline
LOSS           & 0.395 $\pm$ 0.107 & 0.687 $\pm$ 0.178 & 1.295 $\pm$ 0.178 \\
Foundation     & 0.294 $\pm$ 0.077 & 0.523 $\pm$ 0.136 & 1.141 $\pm$ 0.104 \\
CF4 $i$        & 0.429 $\pm$ 0.018 & 0.747 $\pm$ 0.039 & 1.071 $\pm$ 0.016 \\
CF4 W1         & 0.475 $\pm$ 0.019 & 0.817 $\pm$ 0.041 & 1.191 $\pm$ 0.011 \\
\textbf{Joint} & $\bm{0.459 \pm 0.013}$ & $\bm{0.793 \pm 0.035}$ & - \\
\hline
\end{tabular}
\caption{Inferred values of $\beta^\star$, $S_8$ and $b_1$ assuming the linear-theory reconstruction of~\protect\cite{Carrick_2015} for the peculiar velocity samples considered in this work. The ``joint'' sample denotes a combined inference from LOSS, Foundation, and the CF4 $i$- and W1-band \ac{TFR} data. We convert $\beta^\star$ to $S_8$ following the approach of~\cref{sec:growth_rate}. We report the posterior mean and standard deviation. We do not report $b_1$ for the joint sample, as each catalogue has a separate $b_1$.}
\label{tab:C15_beta}
\end{table}

\begin{figure}
    \centering
    \includegraphics[width=\columnwidth]{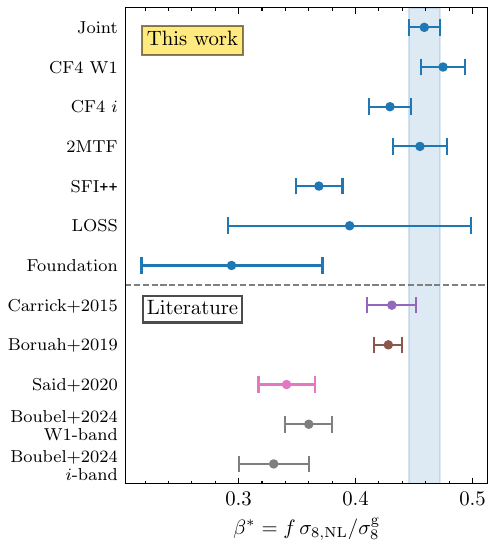}
    \caption{The velocity field scaling parameter $\beta^\star$ of~\protect\citet{Carrick_2015} calibrated against various peculiar velocity samples, compared to previous literature calibration~\protect\citep{Carrick_2015,Boruah_2019,Said_2020,Boubel_2024}. Joint indicates LOSS, Foundation, CF4 TFR W1-band, and CF4 TFR $i$-band. The error bars indicate $1\sigma$.}
    \label{fig:C15_beta}
\end{figure}

Finally, we transform the posterior samples of $\beta^\star$ into $S_8 \equiv \sigma_{8,\mathrm{L}} \sqrt{\Om / 0.3}$. Using the value of $\beta^\star$ calibrated on the joint sample, we find $S_8 = 0.793 \pm 0.035$ under the assumption that $\Om = 0.3111$, based on the mapping of $\sigma_8$ described above. In contrast, applying the mapping of~\cite{Juszkiewicz_2010} yields $S_8 = 0.820 \pm 0.036$ for the same $\beta^\star$. Our inferred $S_8$ is in good agreement with~\cite{Boruah_2019}, while using larger \ac{TFR} samples. In contrast, we find a substantially higher value than~\citet{Said_2020}, who reported $S_8 = 0.637 \pm 0.054$. Assuming $\Om$ values of either $0.25$ or $0.35$ alters our results by less than one per cent. 

In~\cref{fig:C15_S8} we compare the inferred $S_8$ to various values from the literature. Our results are in a good agreement with other peculiar velocity measurements of $S_8$ of~\cite{Boruah_2019,Nusser_2017,Huterer_2017}, but disagree significantly with~\cite{Boubel_2024,Said_2020} who claimed a $>3\sigma$ growth-rate tension with the \ac{CMB}. However, our constraints agree with weak lensing results to within $1\sigma$~\citep{Wright_2025,Carlos_2024,DES_KIDS,DES_Y3} and to within $1.4\sigma$ with \textit{Planck} TT, TE, EE+lowE+lensing~\citep{Planck_2020_cosmo}. \cite{Hollinger_2024} used mock \TWOMPP\ realisations in conjunction with (smaller than \ac{CF4}) mock \ac{TFR} catalogues to show that the sample variance in $f\sigma_{8,\mathrm{NL}}$ is approximately five per cent. Incorporating this additional (conservative) five per cent uncertainty into the $S_8$ constraint would result in $S_8 = 0.793 \pm 0.053$.

\begin{figure}
    \centering
    \includegraphics[width=\columnwidth]{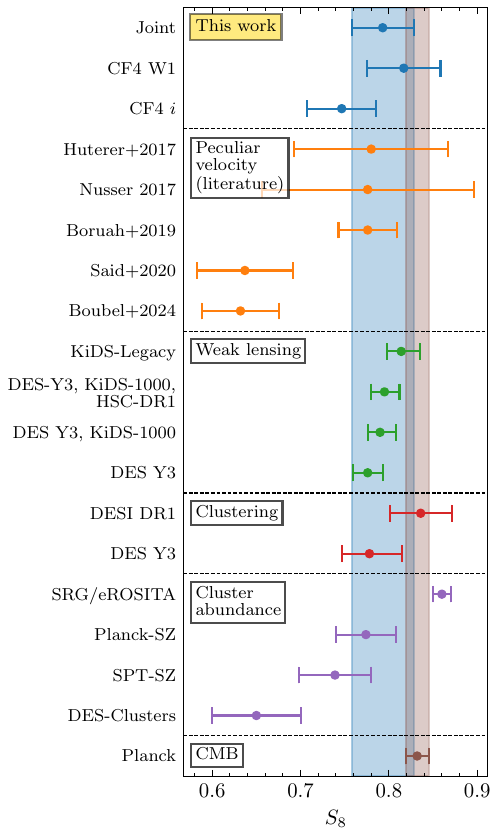}
    \caption{The $S_8 \equiv \sigma_{8,\mathrm{L}} \sqrt{\Om / 0.3}$ parameter inferred with the~\protect\cite{Carrick_2015} linear field calibrated against the joint peculiar velocity sample (LOSS+Foundation+CF4 TFR W1-band+CF4 TFR $i$-band). We compare to literature results using peculiar velocities~\protect\citep{Huterer_2017,Nusser_2017,Boruah_2019,Said_2020}, weak lensing~\protect\citep{Wright_2025,Carlos_2024,DES_KIDS,DES_Y3}, clustering~\protect\citep{DESI_DR1_fullshape,DESY3_clustering}, cluster abundance~\protect\citep{Ghirardini_2024,SPTSZ_clusters,Planck_clusters,DES_clusters} and \textit{Planck} TT, TE, EE + lowE + lensing~\protect\citep{Planck_2020_cosmo}. The errors are $1\sigma$.}
    \label{fig:C15_S8}
\end{figure}


\section{Discussion}\label{sec:discussion}

In this section, we provide a detailed summary of the results (\cref{sec:discussion_summary}), discuss their interpretation (\cref{sec:discussion_interpretation}), and address the caveats of our analysis (\cref{sec:discussion_caveats}). We then compare our results with the literature (\cref{sec:discussion_comparison}) and suggest possible future studies (\cref{sec:future_work}).

\subsection{Summary of results}\label{sec:discussion_summary}

We have evaluated several local Universe reconstructions using various peculiar velocity samples and metrics, ranging from a quantitative to more qualitative analyses. Our first metric is the Bayesian evidence comparison of the reconstructions, shown in~\cref{fig:lnZ_comparison}. This comparison consistently and strongly favoured \CBb\ over the other reconstructions. The second and third ranking reconstructions are those of~\citetalias{Lilow2024} and~\citetalias{Carrick_2015}, respectively, both of which, like \CBb, are constrained using the \TWOMPP\ data. The earlier version of \ac{BORG}, \CBa, performs slightly worse than \CBb, likely stemming from the fact that the \acp{IC} of \CBb\ were constrained using a more accurate, 20-step \texttt{COLA} gravity model, whereas the \acp{IC} of \CBa\ were based on a 10-step particle-mesh gravity model (for further details, see~\citealt{Stopyra_2023}). In contrast, the CosmicFlows-based reconstructions of~\citetalias{Sorce_2018} and~\citetalias{Courtois2023} consistently underperform when tested against both \ac{SN} and \ac{TFR} samples. This outcome is particularly surprising when testing these reconstructions on the samples that were used to constrain them, for which the other models still perform better. We comment on this further in~\cref{sec:discussion_caveats}.

Previous studies typically set $\sigma_v$ to $150~\kmsec$ based on a comparison to $N$-body simulations~\citep{Carrick_2015, Boruah_2019,Said_2020}. However, $\sigma_v$ depends not only on the reconstruction method but also on the peculiar velocity sample. For this reason, we instead treat $\sigma_v$ as a free parameter of the model with a reference (scale-invariant) prior $\pi(\sigma_v) \propto 1 / \sigma_v$. As shown in~\cref{fig:smoothing_effect} and~\cref{fig:sigmav_comparison}, the inferred $\sigma_v$ correlates inversely with the goodness-of-fit and hence $\sigma_v$ provides a good indicator of the quality of a reconstruction.  However, because of the dependence of $\sigma_v$ on the sample in question, it should not be compared between samples, but only between reconstructions for a given sample. For example, a galaxy catalogue with primarily field galaxies will have a lower $\sigma_v$ than a catalogue with galaxies in clusters because of the greater small-scale motions (not captured by the reconstructions) in the latter.

We then examined the parameter $\beta$, which scales the predicted velocity field. For~\citetalias{Carrick_2015}, we denote it as $\beta^\star$ as it is related to the growth rate of structure (see~\cref{sec:growth_rate}). For the other reconstructions, where velocities are consistently sourced by the total matter distribution (except~\citetalias{Lilow2024}), scaling the predicted velocity field with $\beta$ should not be necessary, implying $\beta = 1$. However, even in these cases, if allowed to vary, the recovered $\beta$ can be affected by the resolution at which the field is constrained, inaccurate reconstruction or discrepancies between the cosmological parameters assumed by the reconstruction and the actual description of the Universe. For example, an overly smoothed field may exhibit suppressed velocities, necessitating $\beta > 1$. We tested this hypothesis in~\cref{fig:beta_comparison} by treating $\beta$ as a free parameter for all reconstructions. We find that \CBb\ is consistent with $\beta = 1$ for most samples, except for \ac{CF4} \ac{TFR} $i$-band samples, where it is biased high (though at less than $3\sigma$). Interestingly, both~\citetalias{Sorce_2018} and~\citetalias{Courtois2023} require higher values of $\beta$ on this catalogue as well. On the other hand, both \CBa\ and \citetalias{Lilow2024} prefer a lower $\beta$ value across all samples, suggesting that they overpredict peculiar velocities.

Regarding the external flow magnitude, it is noteworthy that~\citetalias{Lilow2024} indicate their neural network reconstruction is sensitive to ``super-survey'' scales through redshift-space distortions. Our findings in~\cref{fig:Vext_comparison} support this claim, was the external flow magnitude required by~\citetalias{Lilow2024} is nearly consistent with zero, a distinction not seen in other reconstructions which also use the same redshift-space data. The preferred model, \CBb, also favours a relatively small external flow, whereas the less favoured \CBa\ requires a substantially larger flow magnitude. However, similar to $\beta$, there is no clear relationship between the goodness-of-fit and the external flow magnitude.

The \ac{BORG} inferences self-consistently model redshift-space distortions: using the sampled density and velocity field, forward-modelled to $z = 0$, the predicted galaxy counts are mapped into redshift space using the corresponding velocity field and directly compared to the observed redshift-space distribution. However, the finite size of the box used to constrain the \acp{IC} prevents the modelling of the largest-scale power spectrum modes. As a result, some large-scale power may be missing in the redshift-space mapping (or in the density field). In the modelling employed here, we assume an existing reconstruction and forward-model the redshifts from the real-space positions constrained by the direct-distance tracers, thereby self-consistently accounting for redshift-space distortions. To address the missing large-scale power, we include $\bm{V}_{\rm ext}$, representing the contribution to the reconstructed velocity field sourced by external matter.

Lastly, we analysed the predicted bulk flow curves by combining the internal bulk flow of each reconstruction with the external flow. Near the origin, our position provides a constraint on the bulk flow through the \ac{LG}-\ac{CMB} velocity, as derived from the \ac{CMB} dipole~\citep{Planck_2020}. As shown in~\cref{fig:BF_CMB_origin}, all reconstructions (with the possible exception of \CBa) are consistent with this velocity when accounting for $\sigma_v$, which reflects the expected scatter between a given galaxy's velocity and the velocity field estimate at its position.

Regarding the radial dependence, we find that while each reconstruction tracks a somewhat distinct bulk flow curve, there is a general agreement in the bulk flow direction at radii of approximately $100~\Mpch$ among~\citetalias{Carrick_2015}, \citetalias{Sorce_2018}, \citetalias{Courtois2023}, and \citetalias{Lilow2024}. The \ac{BORG} reconstructions, however, exhibit bulk flows that point in different directions, with \CBb\ aligning more closely with those mentioned but slightly shifted towards the Shapley Supercluster (approximately $\ell = 305^\circ$, $b = 32^\circ$). Since our peculiar velocity samples are generally limited to redshifts below $z=0.05$, we do not draw comparisons with~\cite{Watkins_2023}, who report unexpectedly large and increasing bulk flows at radii of $150$ to $250~\Mpch$.

\subsection{Interpretation of results}\label{sec:discussion_interpretation}

While we have determined the relative preferences for the local Universe reconstructions, the reasons underlying these preferences remain unclear. To address this, we extend our analysis in two ways. First, we divide the galaxy samples into subsets based on their observed redshift, inclination errors, or absolute magnitude, and repeat the evidence comparison for each subset. Second, we correlate differences in log-likelihoods between reconstructions with galaxy properties in an attempt to discover which galaxy types drive the preference of some reconstructions over others.

We begin by considering tracers with observed redshift $\zCMB < 0.025$ to examine the inner regions of the reconstructions, compared to the usual threshold of $\zCMB < 0.05$. The evidence ratios for all peculiar velocity samples remain qualitatively identical to those presented in~\cref{fig:lnZ_comparison}: \CBb\ is the most preferred reconstruction,~\citetalias{Lilow2024} and~\citetalias{Carrick_2015} exhibit the second- and third-best goodness-of-fit, respectively, with similar performance, while the CosmicFlows reconstructions are disfavoured. Next, we focus only on the \ac{CF4} \ac{TFR} $i$- and W1-band samples, dividing them into subsets of galaxies with either below- or above-average inclination errors, and analyse the evidence ratios in both cases. As before, the conclusions from our fiducial analysis are robust to these variations. Since inclination errors propagate into the linewidth measurements and are already accounted for, this outcome is unsurprising. It does however indicate that there are no systematic errors associated with the linewidth that scale with the observational uncertainties. Lastly, we estimate the absolute magnitude of each galaxy from its apparent magnitude and redshift-space distance (converting the observed redshift in the \ac{CMB} frame into a distance modulus). The two \ac{CF4} \ac{TFR} samples are then split into subsets of the $25$ per cent brightest and faintest galaxies, and the evidence ratios are recomputed. Yet again, we find that our qualitative conclusions are unchanged.

Next, we repeat our fiducial analysis on the \ac{CF4} \ac{TFR} sample without splitting the data. We record the log-likelihood of each galaxy at each posterior sample and compute the average likelihood $\langle L \rangle$. Subsequently, we calculate the differences in $\log \langle L \rangle$ between reconstructions on the same peculiar velocity samples and correlate these differences with galaxy properties: sky position, observed redshift, apparent magnitude, linewidth, linewidth error, inclination, and inclination error.

Among these features, significant trends emerge only for linewidth and observed redshift. The most pronounced differences occur when comparing \CBb\ with~\citetalias{Courtois2023}. In~\cref{fig:dlogL_eta_CB2_CF4}, we show a scatter plot of the differences in $\log \langle L \rangle$ between \CBb\ and~\citetalias{Courtois2023} as a function of linewidth in the~\ac{CF4} \ac{TFR} W1-band sample. We find a Spearman correlation coefficient of $0.18$ ($p$-value of $10^{-27}$), indicating that \CBb\ is favoured more for galaxies with larger linewidths, corresponding to more massive objects. Since these trace higher-density regions, this result reaffirms that the improved performance of \CBb\ is partly due to its better non-linear modelling via \ac{BORG}. Similar significant trends are observed when comparing \CBb\ with all other reconstructions, including \CBa.

\cref{fig:dlogL_zobs_CB2_CF4} illustrates that \CBb\ consistently achieves higher likelihoods for galaxies at greater distances when compared to the least effective reconstructions~\citetalias{Courtois2023} (Spearman correlation coefficient of $0.23$, $p$-value of $10^{-42}$) or~\citetalias{Sorce_2018}. This suggests that the constraints from CosmicFlows-based reconstructions degrade more rapidly with distance, despite the reconstructed volume in~\citetalias{Courtois2023} extending up to $z=0.08$. However, even though \CBb\ demonstrates better performance for brighter and more distant galaxies, we have confirmed that its preference remains significant even when considering only nearby or faint objects. The trend with observed redshift is not found for either~\citetalias{Carrick_2015} or~\citetalias{Lilow2024} when compared to \CBb. Interestingly, there is no preference between \CBb and~\citetalias{Courtois2023} for the nearest or lowest-linewidth galaxies.

\begin{figure*}
    \centering
    \begin{subfigure}[t]{0.48\textwidth}
        \centering
        \includegraphics[width=\textwidth]{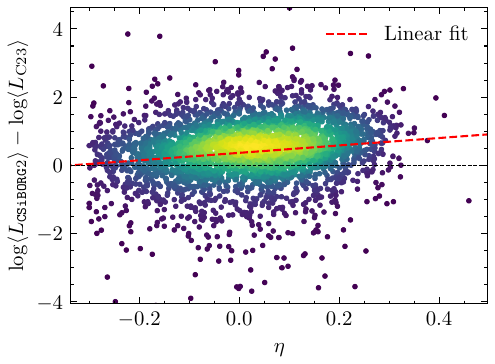}
        \caption{The difference in $\log \langle L\rangle$ as a function of linewidth $\eta$. The Spearman correlation coefficient is $0.18$ ($p$-value of $10^{-27}$).}
        \label{fig:dlogL_eta_CB2_CF4}
    \end{subfigure}
    \hfill
    \begin{subfigure}[t]{0.48\textwidth}
        \centering
        \includegraphics[width=\textwidth]{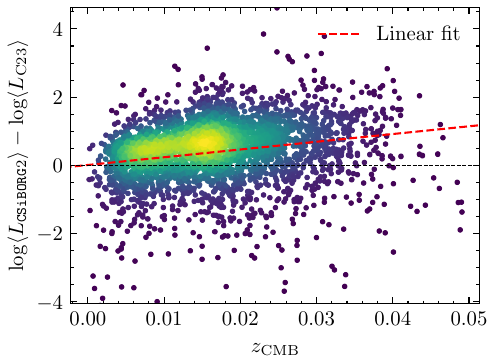}
        \caption{The difference in $\log \langle L\rangle$ as a function of observed redshift $\zCMB$. The Spearman correlation coefficient is $0.23$ ($p$-value of $10^{-42}$).}
        \label{fig:dlogL_zobs_CB2_CF4}
    \end{subfigure}
    \caption{Differences in the logarithms of the average likelihoods of individual galaxies $\log \langle L \rangle$, averaged over the posterior samples, between \CBb\ and the reconstruction by~\protect\cite{Courtois2023}. The plots illustrate these differences as functions of linewidth $\eta$ (left) and observed redshift $\zCMB$ (right). Positive values indicate preference for \CBb and we include a line of best fit. This indicates that \CBb\ is preferentially better than the field of~\protect\cite{Courtois2023} for higher linewidth (mass) galaxies and at higher redshifts. No significant trends are found with other galaxy properties.}
    \label{fig:combined_dlogL}
\end{figure*}

\subsection{Caveats}\label{sec:discussion_caveats}

One caveat in our comparison is that the goodness-of-fit depends on the resolution of the reconstruction. Higher resolution reconstructions would be expected to be preferred simply because they can capture smaller scale motions (before they become noise-dominated). To test this, we smoothed all fields to the same, coarse resolution shown in~\cref{fig:lnZ_comparison}, which did not change our main conclusions: CosmicFlows-based reconstructions remain disfavoured, though~\citetalias{Lilow2024} becomes preferred over \CBb. However, reaching higher resolutions is an advantage of a reconstruction method as, in principle, the goal is to reconstruct the local Universe down to non-linear scales, reducing the required extra random scatter $\sigma_v$. The preference of the \TWOMPP\ derived reconstructions over CosmicFlows-derived ones suggests either that the former dataset is more informative, or that there are limitations of the methods applied to the CosmicFlows data. On one hand, this result is surprising as the peculiar velocities probe directly the matter distribution of the Universe, whereas working with galaxy fields requires a careful bias treatment. On the other hand, while the \TWOMPP\ galaxy redshift-space positions are in practice nearly only shot-noise dominated, the CosmicFlows peculiar velocity data is noisy and sparse: out of the $38,000$ objects in the grouped \ac{CF4} catalogue used by~\citetalias{Courtois2023} only $18,000$ are within $\zobs < 0.05$.

We aimed for a broad comparison of the reconstructions available in the literature, without separating the effect of the reconstruction technique and the input data. For instance,~\citetalias{Sorce_2018} is based on a grouped version of the \ac{CF2} catalogue, which contains approximately only 5,000 galaxies. In contrast,~\citetalias{Courtois2023}, another reconstruction based on peculiar velocities, uses the larger \ac{CF4} catalogue, yet performs similarly to~\citetalias{Sorce_2018} according to our metrics. We demonstrated that the non-linear \ac{BORG}-based reconstructions outperform the other methods. However, this improvement does come at a significantly higher computational cost to obtain the \acp{IC}. Once the \ac{BORG}-based reconstructions are produced, though, this computational cost is irrelevant for any follow-up analysis. In this respect it is worth noting that reconstruction of the velocity field is not a main aim of the \ac{BORG} programme, but rather it comes ``for free.'' Nevertheless, it is worth emphasising that the linear method of~\citetalias{Carrick_2015}, while orders of magnitude less computationally expensive than \ac{BORG}, still performs better than both~\citetalias{Sorce_2018} and~\citetalias{Courtois2023}.

We do not evaluate the quality of the reconstructions in specific regions but instead use all-sky samples. The \ac{TFR} samples mainly include disk (field) galaxies, leading to a focus on lower-density regions rather than galaxy clusters. Evaluating velocities in clusters, which could be done with a fundamental plane sample, is beyond our present scope. However, we discovered in~\cref{sec:discussion_interpretation} that the preference for \ac{BORG} is even greater for brighter galaxies. Since these trace on average higher-density regions, it is likely that the fundamental plane would provide even greater evidence for \ac{BORG}. An example of related work is that by~\cite{Sorce_2024} who examined velocities along the \ac{LOS} to nearby clusters.

A potential systematic relevant for all large-scale structure analyses is the Doppler modulation induced by the observer's motion with respect to the cosmic rest frame (e.g.~\citealt{Ellis_1984,Kaiser_1987,Nusser_2011,BahrKalus_2021}). This effect introduces a dipolar anisotropy in the observed source distribution, and can bias both the inferred density and peculiar velocity fields. In the literature, related systematics are sometimes discussed under the heading of the ``Kaiser-rocket effect,'' which more commonly refers to the dipolar modulation of the two-point correlation function resulting from the observer’s motion~\citep{Kaiser_1987}. While the Kaiser-rocket effect in its original formulation is a concern primarily for correlation function analyses, its underlying physical origin (Doppler boosting) can more generally impact forward-modelling frameworks such as ours if not properly accounted for.

This arises because observer motion induces an apparent overdensity of sources in redshift space along the direction of motion. For example,~\ac{BORG} mitigates this effect by boosting the observed redshifts into the \ac{CMB} frame, thereby adopting the \ac{CMB} frame as the cosmic rest frame. Moreover, since e.g. both~\citetalias{Carrick_2015} and \ac{BORG} self-consistently compute the peculiar velocities, in practice this effect is not a problem. In this work, we also use \ac{CMB}-boosted redshifts and additionally introduce $\bm{V}_{\rm ext}$ as a free parameter to capture any residual effects of the observer's motion (or of the bulk motion of the sample relative to the CMB). A relevant test of this effect was performed by~\cite{Nusser_2014}, who demonstrated using simulations that, without imposing the cosmic rest frame a priori, the motion of the \ac{LG} can only be recovered to within $150$–$200~\kmsec$ in amplitude and $10^\circ$ in direction (assuming linear theory). This is of similar order of magnitude as the behaviour shown in~\cref{fig:BF_CMB_origin}, where we propagate the $\sigma_v$ term (representing the expected redshift scatter between observations and predictions from direct distance tracers and the assumed peculiar velocity), which is typically of order $300~\kmsec$, to the predicted \ac{LG} velocity.

\subsection{Comparison with literature}\label{sec:discussion_comparison}

Peculiar velocity corrections are essential for any local Universe cosmology to model deviations from the Hubble flow. For instance, in the main Pantheon\texttt{+} \ac{SN} analysis~\citep{Brout_2022},~\cite{Carr_2022} define ``Hubble diagram redshifts,'' $z_{\rm HD}$, which are derived by converting the observed heliocentric \ac{SN} (host) redshifts to the \ac{CMB} frame and applying corrections for peculiar motion based on~\citetalias{Carrick_2015}.

The selection of~\citetalias{Carrick_2015} was previously motivated by~\cite{Peterson_2022}, who found it to be the most effective in reducing scatter in the \ac{SN} Hubble diagram. Although the only common reconstruction between~\cite{Peterson_2022} and this work is~\citetalias{Carrick_2015}, their analysis also included a CosmicFlows3-based reconstruction~\citep{Tully_2016}, which was found to increase the scatter rather than reduce it. This conclusion aligns qualitatively with our findings, where we identify the reconstruction by~\citetalias{Carrick_2015} as preferred over the CosmicFlows-based reconstructions. However, we also demonstrate that the latest \ac{BORG} reconstruction has the potential to reduce scatter in the Hubble diagram even further. Moreover, unlike~\citetalias{Carrick_2015}, \ac{BORG} provides a measure of reconstruction uncertainty, allowing for propagation of this uncertainty into the resulting Hubble diagram.

A related study is that of~\cite{Boruah_2021}, who compared~\citetalias{Carrick_2015} to kernel-smoothed velocity field estimates derived from either fundamental plane or \ac{TFR} samples by calculating a Bayesian model evidence. They show that, because of the noise in the velocity estimates going into the kernel-smoothed velocity estimates, the latter are disfavoured when compared to independent peculiar velocity samples.

\subsection{Future work}\label{sec:future_work}

An evident extension of our work is the inclusion of additional peculiar velocity tracers. Two promising candidates are larger \ac{SN} samples and fundamental plane galaxy samples. The primary advantage of \ac{SN} samples lies in their smaller distance errors, approximately 5 per cent, compared to the 20 per cent \ac{TFR} errors. However, \ac{SN} samples are generally smaller in size.

One major \ac{SN} dataset is Pantheon\texttt{+}~\citep{Brout_2022}, which combines several low-redshift \ac{SN} surveys. We excluded this sample from our analysis because it would require careful systematic treatment of the constituent surveys, relying on the covariance matrix provided by~\cite{Brout_2022}. However, this matrix already includes peculiar velocity corrections based on~\citetalias{Carrick_2015}, and disentangling these corrections is non-trivial. Simply subtracting their contributions results in a singular, non-invertible covariance matrix (that parts of the Pantheon\texttt{+} covariance matrix are not positive semi-definite was already noted by~\citealt{Lane_2024}). Another promising \ac{SN} dataset is the Zwicky Transient Facility (ZTF) SN Ia DR2, a volume-limited, unbiased sample of $1000$ \acp{SN} below $\zCMB \sim 0.06$~\citep{Rigault_2024,Amenouche_2024}. However, this dataset is not yet publicly available. Two relevant fundamental plane samples are those from the 6dF Galaxy Survey~\citep{Campbell_2014} and \ac{SDSS}~\citep{Howlett_2022}. Both have been utilised by~\cite{Said_2020} to calibrate the field of~\citetalias{Carrick_2015}, albeit their approach relied on forward modelling the galaxy angular size rather than galaxy distance as we do here. Nevertheless, the consistent evidence comparison across all \ac{SN} and \ac{TFR} samples used in this work indicates that our results are robust to choice of dataset, and hence similar conclusions would likely be reached using Pantheon\texttt{+} or ZTF DR2.


\section{Conclusion}\label{sec:conclusion}

We introduce a novel framework for jointly calibrating distance indicator relations and velocity field reconstructions using peculiar velocity data in the local Universe. Unlike previous studies, we carefully account for uncertainties in distance-indicating properties within a Bayesian hierarchical model. We use reconstructions summarised in~\cref{tab:reconstructions} that include linear theory~\citep{Carrick_2015,Courtois2023}, machine learning~\citep{Lilow2024}, and resimulations of density field constraints derived from Bayesian forward modelling of either redshift-space galaxy positions (\ac{BORG};~\citealt{Jasche_2019,Stopyra_2023}) or peculiar velocities ~\citep{Sorce_2018}. We apply this framework to \ac{TFR} and \ac{SN} samples summarised in~\cref{tab:pv_samples}, with future plans to extend it to fundamental plane samples. We compute the Bayesian model evidence for the joint distance and flow calibration described in~\cref{sec:probabilistic_model}, enabling us to compare how well different local Universe reconstructions explain the available peculiar velocity data.

We show that while by our metrics the older \ac{BORG} reconstruction did not outperform the linear reconstruction of~\citet{Carrick_2015}, the more recent and improved \ac{BORG} reconstruction based on the \acp{IC} of~\cite{Stopyra_2023} is significantly preferred. This suggests a strong future potential for non-linear modelling within the \ac{BORG} framework. We also evaluated the neural-network-based reconstruction of~\cite{Lilow2024}, finding that it performs marginally better than~\cite{Carrick_2015}. Finally, the CosmicFlows-based reconstructions by~\cite{Sorce_2018} and~\cite{Courtois2023} were disfavoured.

We further established that $\sigma_v$, the scatter between predicted and observed redshifts, serves as a reliable indicator of the ability of a reconstruction to explain the data. Moreover, $\beta$, the velocity field scaling parameter, is generally consistent with unity for the preferred \CBb reconstruction (as expected), except in the \ac{CF4} \ac{TFR} $i$-band sample, where it exceeds unity. This suggests that the velocities need to be scaled up to match observations, which is interestingly also true of the two CosmicFlows-based reconstructions.

We also studied the predicted \ac{LG} motion in the \ac{CMB} frame. However, due to the $\sigma_v$ scatter, we found that all reconstructions are broadly consistent with the \ac{CMB}-derived \ac{LG} velocity~\citep{Planck_2020}. We also compared the predicted bulk flow curves as a function of radius and found similar qualitative trends, though the quantitative details vary.

Finally, we calibrated the $\beta^\star$ velocity scaling of the linear theory reconstruction of~\citet{Carrick_2015} using our peculiar velocity samples, enabling us to place a stringent constraint on the parameter $S_8 \equiv \sigma_8 \sqrt{\Om / 0.3}$. For the joint sample, we determined $\beta^\star = 0.459 \pm 0.013$, which translates to $S_8 = 0.793 \pm 0.035$. This estimate of $S_8$ agrees well with both weak lensing results and \textit{Planck}. It also matches some peculiar velocity measurements, but disagrees with others~\citep{Boubel_2024,Said_2020}.

In conclusion, this work provides a comprehensive comparison of various velocity field reconstructions available in the literature, identifying the latest \ac{BORG} reconstruction based on \acp{IC} from~\cite{Stopyra_2023} as the most strongly preferred. Our framework and results will be valuable not only for evaluating local Universe reconstructions but also for cosmological analyses that rely on peculiar velocity corrections, such as \acp{SN} or the dark siren method for gravitational waves as well as growth of structure measurements from peculiar velocities. This is highly relevant in the context of the Hubble tension and, more broadly, for any inference requiring physical distances to galaxies.

\section{Data availability}

The~\cite{Carrick_2015} reconstruction is available at \url{https://cosmicflows.iap.fr}, the \cite{Lilow2024} reconstruction at \url{https://github.com/rlilow/2MRS-NeuralNet} and the~\cite{Courtois2023} reconstruction at~\url{https://projets.ip2i.in2p3.fr/cosmicflows/}. The public \ac{CF4} data is available at \url{https://edd.ifa.hawaii.edu/dfirst.php}. The code underlying this article is available at \url{https://github.com/Richard-Sti/CANDEL} and all other data will be made available on reasonable request to the authors.

\section*{Acknowledgements}

We thank Indranil Banik, Supranta S. Boruah, Pedro G. Ferreira, Sebastian von Hausegger, Wojciech Hellwing, Amber Hollinger, Jens Jasche, Ehsan Kourkchi, Sergij Mazurenko, Stuart McAlpine, Alicja Polanska, Jenny Sorce, Stephen Stopyra, Maria Vincenzi, and Tariq Yasin for useful inputs and discussions. We also thank Jonathan Patterson for smoothly running the Glamdring Cluster hosted by the University of Oxford, where the data processing was performed. This work was done within the Aquila Consortium\footnote{\url{https://aquila-consortium.org}}. The authors would like to acknowledge the use of the University of Oxford Advanced Research Computing (ARC) facility in carrying out this work.\footnote{\url{https://doi.org/10.5281/zenodo.22558}}

RS acknowledges financial support from STFC Grant No. ST/X508664/1, the Snell Exhibition of Balliol College, Oxford, and the CCA Pre-doctoral Program. HD is supported by a Royal Society University Research Fellowship (grant no. 211046). DJB is supported by the Simons Collaboration on ``Learning the Universe.'' This project has received funding from the European Research Council (ERC) under the European Union's Horizon 2020 research and innovation programme (grant agreement No 693024). This work was performed using the DiRAC Data Intensive service at Leicester, operated by the University of Leicester IT Services, which forms part of the STFC DiRAC HPC Facility (www.dirac.ac.uk). The equipment was funded by BEIS capital funding via STFC capital grants ST/K000373/1 and ST/R002363/1 and STFC DiRAC Operations grant ST/R001014/1. DiRAC is part of the National e-Infrastructure. For the purpose of open access, we have applied a Creative Commons Attribution (CC BY) licence to any Author Accepted Manuscript version arising.

\bibliographystyle{mnras}
\bibliography{ref}

\appendix

\section{SN flow model}\label{sec:supernova_flow}

In~\cref{sec:probabilistic_model}, we derived the model for joint density and velocity field calibration, alongside the calibration of the \ac{TFR}. We now extend this framework for \ac{SN} calibration, aiming to infer the Tripp parameters $\mathcal{A}$ and $\mathcal{B}$, as well as the absolute magnitude of Type Ia \acp{SN}, $\MSN$. For this purpose, we consider the observed apparent magnitude $\mobs$, the ``observed'' SALT2 light curve parameters: stretch $\xoneobs$ and colour correction $\cobs$, along with their corresponding ``true'' values $\xonetrue$ and $\ctrue$.

The first difference compared to the \ac{TFR} model is that we assume all sources to have the same absolute magnitude $\MSN$ and adopt the distance–magnitude relation
\begin{equation}
    m_{\rm standard} = \mu(r) + \MSN,
\end{equation}
where $m_{\rm standard}$ is the standardised apparent magnitude.  
Following the Tripp parametrisation in~\cref{eq:tripp}, the predicted \ac{SN} apparent magnitude is
\begin{equation}\label{eq:SN_predicted_magnitude}
    \mpred = \mu(r) + \MSN - \mathcal{A}\,\xonetrue + \mathcal{B}\,\ctrue,
\end{equation}
where we explicitly use the true stretch and colour corrections. In this case there are three latent parameters per source: $r$, $\xonetrue$, and $\ctrue$. We sample $r$ from the same prior as in~\cref{eq:distance_prior}, and place a (correlated) Gaussian prior on $\xonetrue$ and $\ctrue$,
\begin{equation}
   \pi(\xonetrue,\, \ctrue \mid \bm{\theta})
   =
   \mathcal{N}\!\left(\{\xonetrue,\, \ctrue\};\, \bm{\mu}_{\rm SN},\, \mathbf{C}_{\rm SN}\right),
\end{equation}  
where $\bm{\mu}_{\rm SN}$ is the hyperprior mean of $\xonetrue$ and $\ctrue$, inferred with uniform priors. The $2\times 2$ covariance matrix $\mathbf{C}_{\rm SN}$ has standard deviations $\sigma_i$, inferred with the reference (scale-invariant) prior $\pi(\sigma_i) \propto 1/\sigma_i$, and a correlation coefficient assigned a uniform prior in the range $[-1,\,1]$.

Unlike the \ac{TFR} case, where we numerically marginalise over both $r$ and $\etatrue$ at each \ac{MCMC} step, here we marginalise only over $r$ and explicitly sample $\xonetrue$ and $\ctrue$ with \ac{HMC} to avoid evaluating a three-dimensional numerical integral at every step. This results in a high-dimensional posterior, which renders evidence calculation infeasible. For evidence estimation, we therefore approximate the posterior by fixing $\xonetrue$ and $\ctrue$ to their observed values, as described in~\cref{sec:evidence_calculation}. This approach is implicitly followed by~\cite{Boruah_2019} and other earlier works. The rationale for this method is that the observational uncertainties are generally subdominant to the intrinsic scatter. Here, we directly compare the two methods to verify that the approximate method is reliable. \cref{fig:full_vs_approximate_posterior} shows the posteriors from the two methods for the Foundation \ac{SN} sample and the \citetalias{Carrick_2015} reconstruction. We find that the only parameter significantly affected by use of the approximate method is $\mathcal{B}$. The remaining parameters remain unaffected.

Due to the high dimensionality of the posterior, it is not straightforward to directly assess how this approximation impacts the evidence ratios. We instead compute the \ac{BIC}, defined as  
\begin{equation}
    \mathrm{BIC} = D \log N_{\rm gal} - 2 \log \hat{\mathcal{L}},
\end{equation}  
where $D$ is the dimensionality of the posterior, $N_{\rm gal}$ is the number of galaxies, and $\hat{\mathcal{L}}$ is the maximum-likelihood value. The \ac{BIC} can serve as a model selection criterion analogous to Bayesian evidence, where lower \ac{BIC} values indicate a preferred model. We note that the \ac{BIC} is a reliable model selection indicator only when the number of observations is significantly larger than the dimensionality of the model. Nonetheless, we compute the \ac{BIC} values for the full posterior and find good qualitative agreement with our main results presented in~\cref{fig:lnZ_comparison}, which were obtained using the approximated posterior. The ranking of the models remains unchanged, and the magnitude of the differences between them is approximately preserved. Yet another way to test this approximation is to assume a Gaussian posterior and use the Laplace approximation. However, we do not pursue this additional test here, since for the more informative \ac{TFR} we explicitly marginalise over the per-galaxy latent variables instead, which involves no approximation.

The advantage of this framework is that corrections to the Tripp formula can be incorporated directly by modifying~\cref{eq:SN_predicted_magnitude} and introducing additional terms that may depend, for example, on host-galaxy mass or environment. In the present analysis we assume independence between \acp{SN} by omitting any covariance matrix, since the Tripp calibration is inferred jointly, and we neglect correlations between the SALT2 light curve parameters. Additional covariance can, however, be included in the likelihood of either the observed redshift or apparent magnitude. While this would render the one-dimensional marginalisation over radial distance computationally prohibitive, the distances could instead be sampled explicitly. Incorporating more general selection effects in this context is non-trivial and may require a simulation-based approach~\citep{Boyd}.

\begin{figure*}
    \centering
    \includegraphics[width=\textwidth]{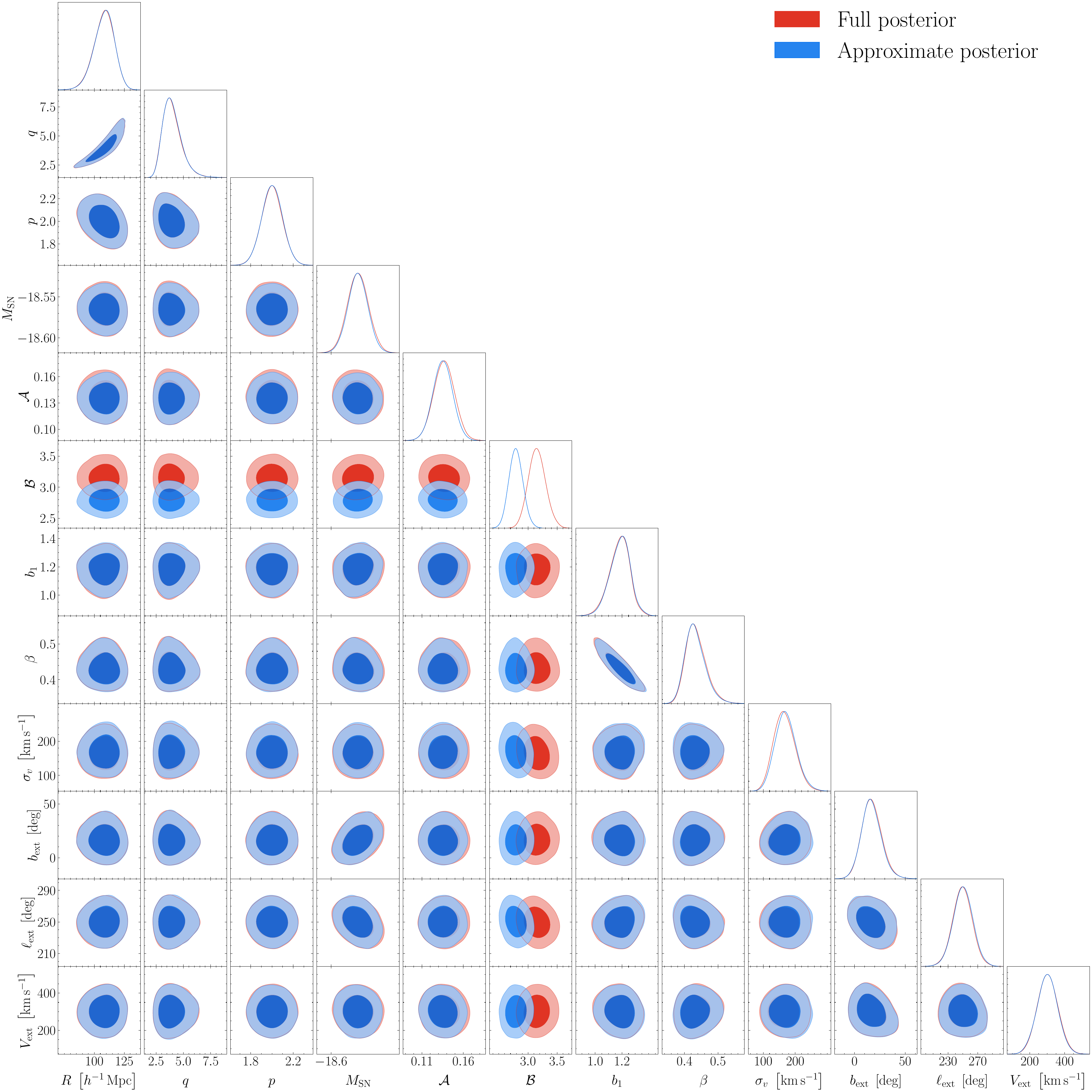}
    \caption{Example comparison of the inferred posteriors on the model parameters using the full \ac{SN} likelihood from~\cref{eq:full_likelihood} and~\cref{sec:supernova_flow} (red) versus the approximate likelihood used for evidence calculation (\cref{sec:evidence_calculation}; blue). The underlying reconstruction is~\protect\cite{Carrick_2015}, and the peculiar velocity sample is the Foundation \ac{SN} sample. The posteriors are nearly identical except for $\mathcal{B}$. Contours represent the $1\sigma$ and $2\sigma$ credible regions.}
    \label{fig:full_vs_approximate_posterior}
\end{figure*}

\bsp
\label{lastpage}\end{document}